\documentclass[final,1p,11pt]{elsarticle}

\usepackage{amsmath,amsfonts, color,bm, lineno,array,multirow,natbib,float,hyperref,todonotes,amssymb,setspace,algorithmicx,algorithm,algpseudocode}

\setcitestyle{authoryear,citesep={;}}

\DeclareMathOperator*{\argmin}{argmin}
	
\journal{}

\newcommand{\norm}[1]{\left\lvert#1\right\rvert}

\begin{document}

\begin{frontmatter}



\title{Multiplicative non-Gaussian model error estimation in data assimilation.}


\author{S. Pathiraja$^1$, P.J. Van Leeuwen$^{2,3}$}

\address{$^1$Institut f{\"u}r Mathematik, Universit{\"a}t Potsdam, Germany\\ $^2$Department of Atmospheric Science, Colorado State University, USA\\
$^3$Department of Meteorology, University of Reading, UK}

\begin{abstract}
Model uncertainty quantification is an essential component of effective data assimilation.  Model errors associated with sub-grid scale processes are often represented through stochastic parameterizations of the unresolved process.  Many existing Stochastic Parameterization schemes are only applicable when knowledge of the true sub-grid scale process or full observations of the coarse scale process are available, which is typically not the case in real applications.  We present a methodology for estimating the statistics of sub-grid scale processes for the more realistic case that only partial observations of the coarse scale process are available.  Model error realizations are estimated over a training period by minimizing their conditional sum of squared deviations given some informative covariates (e.g. state of the system), constrained by available observations and assuming that the observation errors are smaller than the model errors. From these realizations a conditional probability distribution of additive model errors given these covariates is obtained, allowing for complex non-Gaussian error structures. Random draws from this density are then used in actual ensemble data assimilation experiments. We demonstrate the efficacy of the approach through numerical experiments with the multi-scale Lorenz 96 system using both small and large time scale separations between slow (coarse scale) and fast (fine scale) variables. The resulting error estimates and forecasts obtained with this new method are superior to those from two existing methods.\\
\end{abstract}


\end{frontmatter}


\section{Introduction}
\label{S:1}
Model uncertainty quantification is one of the central challenges in successfully utilizing any data assimilation method; the optimal combination of forecasts and measurements is critically dependent on the uncertainties assigned to each.  Model errors can arise from a range of sources, including but not limited to: model discretization errors in space and time, unresolved sub-grid processes, and uncertainties in model forcing or input data.  The lack of complete high resolution and high quality verification data makes model error estimation difficult in most real world applications. \\

Early methods for characterizing model error involved estimating the size and spatial structure of the missing physics \citep{VanLeeuwen1999, VanLeeuwen2001}. In ensemble data assimilation, inflation and localization techniques are used which involve modifying the sample covariance  \citep[e.g.][]{Anderson1999,Hamill2001, Houtekamer2001}.  These were initially developed as a heuristic remedy for filter divergence in ensemble Kalman filtering, but the increase in forecast variance associated with inflation also has the added benefit of at least partially accounting for forecast model errors.  In additive inflation \citep{Anderson1999} the diagonal of the forecast covariance matrix is increased by some additive term $\lambda > 0$ whilst in multiplicative inflation \citep{Anderson2001}, all elements of the covariance matrix are multiplied by a $\lambda > 1$.  
The inflation parameter can either be manually tuned, or more objective adaptive inflation factor estimation can be used \citep{Anderson2007, Miyoshi2011, Liang2012}. 
A more explicit treatment of model error involves estimating a forecast bias term, which can be considered as stochastic or deterministic.  This can be estimated from the difference in mean analyses and forecasts \citep[e.g.][]{Saha1992, Dee1995}, using the difference between 2 forecast models of differing resolution \citep{Hamill2005} or online within the data assimilation system by incorporating a constant additive term to be updated alongside the system states \citep[e.g.][]{Dee1998}.  However, these methods are limited in that the focus is only on the first two moments of the model error distribution.

More recently, there has been renewed interest in off-line estimation of model error statistics using analysis increments (i.e. difference between forecast and analysis) from a data assimilation run \citep{Rodwell2007, Mitchell2015}.  These approaches rely heavily on a Gaussianity assumption.  
Another line of research is in time-varying model error estimation \citet{Brasseur2005} (for a marine biogeochemistry model) or in time varying parameter estimation \citet{Pathiraja2018, Pathiraja_insights2018} which is particularly useful when one knows apriori that a model parameter is non-constant (e.g. land cover in a hydrologic model).  However these approaches cannot capture more general model structural errors.  

In the Variational Data Assimilation literature, model error is often considered by formulating the forecast model as a weak constraint in the optimization problem (often referred to as Weak constraint 4D-Var) \citep{Zupanski1997, Tremolet2006}.  One approach to achieve this is through Long-window 4D-Var, where an additive model error term is incorporated as a control variable in the 4D-Var formulation, with initial conditions for each window held fixed \citep{Fisher2005, Tremolet2006, Fisher2011}.  These approaches require apriori specification of the model error covariance matrix, while our task is to estimate the characteristics of the model error. \cite{Zhu2017} estimated the model error covariance online in a particle filter in a 1000-dimensional Lorenz 1996 model. The advantage of a particle filter is, similar to long-window 4d-Var, that the error covariance of the state plays no role in the estimation. The method needs a first guess of the model error covariance matrix, which is then updated over time, with the restriction that only an additive second order moment is estimated.

More complex methods involve estimating a parameterization on-line using linear regression on a pre-defined large set of potential functional forms \citep[e.g.][]{Lang2016}. Furthermore, off-line methods from machine learning,  such as Relevant Vector Machines \citep{Bishop2006} and Bayesian Symbolic Regression \citep{Jin2020} have been explored to find structural model errors, and hence missing physics. These methods define a set of basic functions and build model equations from these that fit the data best. All these methods have in common that they do need to define a set of functional relations, after which the fitting is performed, which limits the freedom of structure of model errors.  \citet{Bonavita2020} provide an in-depth investigation of how best to integrate machine learning for model error estimation into existing data assimilation methods.  

A related line of research involves stochastic parameterization and model reduction methods to account for model errors associated with unresolved sub-grid scale processes in data assimilation \citep[][e.g]{Mitchell2012, Berry2014, Mitchell2015, Lu2017}.  This is particularly relevant in weather and climate modelling where the system dynamics evolve on a wide range of spatial and temporal scales.  Such model reduction methods involve modelling or parameterizing sub-grid scale processes in a more computationally tractable fashion than solving the true sub-grid differential equations.  Often this is achieved through either deterministic or stochastic parameterizations which aim to capture the mean effects of small scale processes on the resolved variables.  Several studies have demonstrated the superiority of stochastic over purely deterministic parameterizations in this regard \citep{Buizza1999, Palmer2001}.  

Methods for stochastic parameterizations of multi-scale systems vary widely; from homogenization methods that are suited to systems with large time scale separations \citep{Pavliotis2008, Wouters2016} to fitting stochastic models to sub-grid tendencies \citep[][e.g]{Wilks2005, Arnold2013}.  Methods of the form of the latter include that of \citet{Crommelin2008}, who proposed utilising a Conditional Markov Chain to represent the evolution of sub-grid tendencies given the state of the resolved variable.  The transition matrices are estimated using realizations of the true sub-grid tendencies. \citet{kwasniok2012} explored a similar approach whereby a clustering algorithm was used to develop a cluster-weighted Markov chain to represent the sub-grid tendencies.  \citet{Arnold2013} extended the work of \citet{Wilks2005} by examining the potential of autoregressive error models to effectively parameterize sub-grid tendencies in the multi-scale Lorenz 96 system.  This was further extended in \cite{Gagne2020} where a Generative Adversarial Network (GAN) was trained on data of sub-grid tendencies and coarse scale variables from the 2 scale Lorenz 96 model.  \citet{Lu2017} have proposed using a non-Markovian non-linear autoregressive moving average model to characterize model error.  All of the aforementioned approaches require knowledge of the sub-grid scale equations, representative data of the sub-grid tendencies and/or full observations of the resolved variables.  This reduces their applicability for more realistic data assimilation applications where knowledge of the sub-grid scale processes is unavailable and the resolved variables are only partially observed.    

We propose a methodology for model uncertainty estimation that
is specifically designed for partially observed systems and does not require knowledge of the sub-grid scale processes.  The method is suited to systems where a locality and homogeneity 
assumption can be invoked, as this is used to regularize the ill-posed problem of estimating model errors from partial observations. In such systems, errors due to sub-grid scale processes are dependent only on neighboring states instead of the full resolved state vector, and the error statistics are the same at each location in space, or over larger parts of state space with similar physics.  

The approach first approximates the conditional probability distribution of additive model errors given some informative covariates (e.g. state of the system).  This density is calculated from estimated model error realisations.  These are obtained during a training phase by minimizing their variance conditioned on the informative covariates, constrained by available observations.  Samples from the estimated distribution can then by combined with forward model simulations to generate a forecast distribution in any ensemble data assimilation framework. The distribution estimate is nonparametric, allowing for the characterization of highly non-Gaussian errors.  We demonstrate its efficacy through numerical experiments with the multi-scale Lorenz 96 system.   The forecast model in the assimilation experiment is the single layer Lorenz 96, so that model errors arise from the unresolved high frequency fast variables.  The proposed approach is compared to two benchmark methods in terms of the ability to recover the true model error structure and the impact on assimilation and forecast quality. 

The remainder of this paper is structured as follows.  In Section \ref{S:2}, we discuss data assimilation methods and the Ensemble Transform Kalman Filter (ETKF), which is adopted as the assimilation algorithm in this study.  Methods for estimating model uncertainty in partially observed systems are discussed in Section \ref{S:3}.  The details of the proposed method are provided, along with a long window 4D-Var formulation and an ensemble analysis increment based method, both of which are adopted as benchmarks.  In Section \ref{S:4} we describe the numerical experiments with the multi-scale Lorenz 96' system.  We conclude with a summary of the main outcomes and possibilities for future work in Section \ref{S:5}.

\section{Data Assimilation Methods}
\label{S:2}

The general problem setting considered in this study is described as follows.  Suppose the system of interest can be represented by the following discrete time continuous state space equation:
\begin{linenomath}
\begin{equation}
\label{eq:2a}
\bm{x}_j = M(\bm{x}_{j-1}) + \bm{\eta}_j
\end{equation}
\end{linenomath}
where $\bm{x}_{j-1} \in \mathbb{R}^{N_x}$ is the true state vector at time $j-1$; $M: \mathbb{R}^{N_x} \rightarrow \mathbb{R}^{N_x}$ is a Markov Order 1 forecast model; and $\bm{\eta}_j \in \mathbb{R}^{N_x}$ is an additive model error at time $j$ capturing deficiencies in the forecast model $M$. \\

Noisy partial observations of the state $\bm{x}_j$ are available, given by the following:
\begin{linenomath}
\begin{equation}
\label{eq:2b}
\bm{y}_j = \mathbf{H}\bm{x}_j + \bm{\varepsilon}_j
\end{equation}
\end{linenomath}
where $\mathbf{H}:\mathbb{R}^{N_x} \rightarrow \mathbb{R}^{N_y}$ is a $ N_y \times {N_x} $ matrix consisting of 1's and 0's only (i.e. state components are either directly observed or not at all), $\bm{y}_j \in \mathbb{R}^{N_y}$ is the vector of observations at time $j$, and $\bm{\varepsilon}_j \in \mathbb{R}^{N_y}$ is the observation noise at time $j$, assumed to be temporally uncorrelated Gaussian with zero mean and known covariance matrix $\mathbf{R} \in \mathbb{R}^{N_y \times N_y}$.  In this study, we focus on the case $ N_y < {N_x}$, i.e. the state vector is partially observed.  These observations are available at a coarser temporal resolution than the model forecast time step.  Throughout the manuscript, the notation $\bm{v}[k]$ is used to refer to the $k$th element of some vector $\bm{v}$; $\mathbf{A}[k,l]$ refers to the element at the $k$th row and $l$th column of some matrix $\mathbf{A}$. \\

The aim of data assimilation is to optimally combine observations and prior information (usually from a numerical model, e.g. (\ref{eq:2a})) based on their respective uncertainties.  The standard discrete time Kalman filter provides the optimal posterior (in the minimum variance sense) for the special case of linear forecast model and observation operator, and for zero mean temporally uncorrelated Gaussian process and observation noise.  Ensemble Kalman Filter methods (amongst others) have been developed for high-dimensional systems where the full covariance matrix is too large to store in a computer, with an additional benefit for the more general case of non-linear and non-Gaussian problems encountered in many applications.  The Ensemble Transform Kalman Filter (ETKF) \citep{Bishop2001, Wang2004} has been widely adopted particularly in meteorological data assimilation due to its computational efficiency and accuracy in high dimensional systems with small ensemble sizes when localization is applied. 

The ETKF is an extension of the original ensemble Kalman filter (EnKF) proposed by \citet{Evensen1994}.  It belongs to the class of ensemble square root filters which operate on the square root of the forecast and analysis error covariance rather than the full covariance matrices \citep{Tippett2003, Vetra2018}.  Such methods use a deterministic transformation to map the forecast ensemble to the analysis ensemble, whose statistics are consistent with the Kalman filter update. As noted by \citet{Tippett2003}, the linear transformation is not uniquely defined, and is the main distinguishing factor between different ensemble square root methods.  Here we present the method of \citet{Wang2004}, which is an updated version of the original ETKF proposed by \citet{Bishop2001} that ensures the filter is unbiased.  A single cycle of the ETKF is summarized below.\\

A forecast ensemble at time $j$ (denoted $\mathbf{X}_j^f$) is generated by propagating the analysis ensemble from the previous time through (\ref{eq:2c}): 
\begin{linenomath}
\begin{gather}
\label{eq:2c}
\bm{x}_j^{f,i} = M(\bm{x}_{j-1}^{a,i}) + \bm{\eta}_j^{i} \quad \forall \quad i \in \{1, ..., n\} \\
\label{eq:2d}
\mathbf{X}_j^f = \begin{bmatrix} \bm{x}_j^{f,1},&... &,\bm{x}_j^{f,n} \end{bmatrix} \: \in \mathbb{R}^{{N_x} \times n}  
\end{gather}
\end{linenomath}
where the superscripts $f$ and $a$ denote the forecast and analysis, respectively. The crucial strength of ensemble Kalman filters is that one can avoid the explicit calculation of the state covariance matrices.  In the ETKF, this is achieved by writing the analysis ensemble deviation matrix $\mathbf{X}_j^{a'}$ in terms of the forecast ensemble deviation matrix $\mathbf{X}_j^{f'}$ as
\begin{linenomath}
	\begin{equation}
	\label{eq:2j}
	\mathbf{X}_j^{a'} = \mathbf{X}_j^{f'} \mathbf{T} \\
	\end{equation}
\end{linenomath}
where $\mathbf{X}_j^{f'} := \mathbf{X}_j^f - \overline{\bm{x}}_j^{f}\bm{k}^T \; \in \mathbb{R}^{{N_x} \times n} $, $\overline{\bm{x}}_j^{f}$ is the ensemble mean, $\bm{k}$ is a vector of ones and 
$\mathbf{T} \in \mathbb{R}^{n \times n}$ is a transformation matrix given by 
\begin{linenomath}
	\begin{equation}
	\label{eq:2o}
	\mathbf{T} = \mathbf{U}(\mathbf{I} + \mathbf{\Sigma} \mathbf{\Sigma}^T)^{-1/2}\mathbf{U}^T 
	\end{equation}
\end{linenomath}
where $\mathbf{U}$ and $\mathbf{\Sigma}$ arise from the SVD of the scaled forecast ensemble observation deviation matrix $\mathbf{W}$, i.e. 
\begin{linenomath}
	\begin{align}
	\label{eq:2n}
	\mathbf{W} := \frac{1}{\sqrt{n-1}} \left([\mathbf{X}_j^{f'} ]^T \mathbf{H}^T \mathbf{R}^{-1/2}\right) 
	 = \mathbf{U}\mathbf{\Sigma}\mathbf{V}^T.  
	\end{align}
\end{linenomath}
 This approach transforms the computations to ensemble space which significantly reduces the required number of operations whenever $n << N_y$ (as is typically the case in real world geophysical applications).  Finally, the SVD of $\mathbf{W}$ is utilized to efficiently calculate the analysis ensemble mean $\overline{\bm{x}}_j^{a}$: 
\begin{linenomath}
\begin{equation}
\label{eq:2p}
\overline{\bm{x}}_j^{a} = \overline{\bm{x}}_j^{f} + \frac{1}{\sqrt{n-1}} \mathbf{X}_j^{f'} \mathbf{U} \left(\mathbf{\Sigma}^{T} \mathbf{\Sigma} + \mathbf{I}\right)^{-1} \mathbf{\Sigma} \mathbf{V}^T \mathbf{R}^{-1/2} (\bm{y}_j - \mathbf{H}\overline{\bm{x}}_j^{f}) 
\end{equation}
\end{linenomath}

\section{Accounting for Model Uncertainty}
\label{S:3}
In the following, we propose a method for estimating model uncertainty in partially observed systems where knowledge of the unresolved processes is unavailable.  The approach is specifically for use with Monte-Carlo based sequential filtering techniques such as Ensemble Kalman methods and Particle methods.  Existing methods for accounting for model errors that are amenable to the partially observed setting are also discussed in Section \ref{S:32}.

\subsection{Proposed Method}
\label{S:31}
The proposed method utilizes a training period to obtain estimates of model errors using an optimization procedure and knowledge of some informative covariates (e.g. state of the system).  Estimates are generated by
minimizing the conditional sum of squared deviations of the model errors given the covariates, constrained by available observations.  These estimates are then used to build a conditional model error probability density using kernel density estimation, which allows for the characterization of potentially non-Gaussian features. In actual data assimilation experiments model errors are drawn from this conditional distribution. Since the density estimation is computed off-line the cost of incorporating uncertainty in this fashion is kept to a minimum.

The following assumptions are required for the method:
\begin{enumerate}
	\item The system states are directly but partially observed, i.e. $\mathbf{H}$ takes the form as described in Section \ref{S:2}.
	\item The additive error at time $j$ and grid point $k$, $\bm{\eta}_j[k]$, is dependent on a vector of informative covariates denoted by $\bm{z}_{j,k}$.  For instance, if it is assumed that the error depends only on the state at the previous time and same location, then $\bm{z}_{j,k} = \bm{x}_{j-1}[k]$.  Other possibilities include $\bm{z}_{j,k} = \left[\bm{x}_{j-1}[k], \bm{x}_{j-1}[k-1], \bm{x}_j[k+1] \right]$ if the error is expected to depend on the states in a neighbourhood of the grid point, and $\bm{z}_{j,k} = \left[\bm{x}_{j-1}[k], \bm{x}_{j-2}[k] \right]$ if a longer temporal dependence on the states is expected.    
	\item The magnitude of the measurement errors is small in comparison to the magnitude of the model errors, i.e. $\lVert \bm{\varepsilon}_j \rVert << \lVert \bm{\eta}_j \rVert $
	\item Additive error statistics are the same in time and space, i.e.
	$p(\bm{\eta}_j [k] | \bm{z}_{j,k}) \equiv p(\bm{\eta}_m [l] | \bm{z}_{m,l}) \enskip \forall \enskip l,k \in \{1, 2, \cdots, K\}, \enskip j,m \in \{1, 2, \cdots, T\}$.
\end{enumerate}

The key advantages of the proposed approach are 1) it allows for the estimation of complex error structures with minimal apriori knowledge and partial observations; 2) requires no assumptions or specification of a parametric error distribution (e.g. Gaussian errors) and considers the full range of moments (not just bias and covariance); 3) computes all error statistics from data, without the need for numerical tuning; and 4) has sufficient flexibility to incorporate a range of covariates that influence error processes, which will generally be problem dependent.  

The methodology consists of two main steps and is discussed in detail for the remainder of this section.  Unless otherwise stated, the notation $\bm{v}_{t_1:t_2}$ is used to indicate the sequence of vectors 
$\{\bm{v}_j\}_{j = t_1, t_1+1, \cdots, t_2} $.  The hat notation  $ \hat{} $  is used to indicate an estimate of a variable.  We also let $k_u$ and $k_o$ denote the vector of indices of the unobserved and observed grid points, so that for instance $\bm{\eta}_j[k_u]$ denotes the vector of errors at time $j$ at the unobserved grid points.

\subsubsection{Step 1. Offline additive error estimation}
\label{sec:step1}

Given a training period of length $T$ time steps, the aim is to estimate the sequence of errors $\bm{\eta}_{1:T}$ under the assumptions stated above. To this end we solve a constrained optimization problem where the objective function is of conditional sum of squares type: 
\begin{linenomath}
\begin{align}
\label{eq:new14}
\hat{\bm{\eta}}_{1:T} = \argmin_{\bm{\eta}_{1:T}} \sum_{k=1}^K \sum_{j = 1}^T \left(\bm{\eta}_j[k] - \hat{m}(\bm{z}_{j,k}) \right)^2 
\end{align}
\end{linenomath}
subject to the constraints
\begin{linenomath}
\begin{align*}
    \bm{y}_j &= \mathbf{H} \bm{x}_j \quad \forall \enskip j = 1,2, \cdots T \\
    \bm{x}_j &= M(\bm{x}_{j-1}) + \bm{\eta}_j.
\end{align*}
\end{linenomath}
$\hat{m}(\bm{z}_{j,k})$ is the Nadaraya–Watson Kernel estimator of 
$\mathbb{E}(\bm{\eta}_j[k]| \bm{z}_{j,k})$, given by 
\begin{linenomath}
\begin{align}
\label{eq:condmean}
\hat{m}(\bm{z}_{j,k}) := \frac{\sum_{l=1}^K \sum_{i=1}^T K_b(\norm{\bm{z}_{i,l} - \bm{z}_{j,k}}) \bm{\eta}_i[l]}{\sum_{l=1}^K \sum_{i=1}^T K_b(\norm{\bm{z}_{i,l} - \bm{z}_{j,k}})}
\end{align}
\end{linenomath}
where $K_b$ is a kernel function with bandwidth $b$, both of which must be selected.  Common choices for the kernel function could be a Gaussian, Uniform or Epanechnikov kernel.  Such regularizers are also used in semi-supervised learning where they guide the learning method to find models that respect some underlying structure of the samples. The Levenberg-Marquardt algorithm is used as the minimizer.

Optimizing $\bm{\eta}_{1:T}$ can be prohibitively expensive especially for large $T$ and ${N_x}$.  We therefore use a sequential optimization technique over a sliding time window of length $\tau$, as is also employed in Long window weak-constraint 4D-Var \citep{Tremolet2006} and particle smoothing methods \citep{Sarkka2013}.  For a given time $t$, initial condition estimate $\hat{\bm{x}}_{t-1}$ and time window length $\tau$, the optimization problem (\ref{eq:new14})-(\ref{eq:condmean}) is restricted to $j \in \{t, t+1, \cdots, t+\tau \}$ instead of $j \in \{1, 2, \cdots, T\}$.  This process is then repeated by sliding the window of length $\tau$ forward one time step, so that  ${\bm{\eta}}_{t+1:t+\tau+1}$ is optimized, where the existing estimates from the previous optimization step are used as an initial guess.   The sliding window procedure allows one to avoid specifying the background error covariance matrix or including the initial condition $\hat{\bm{x}}_{t-1}$ in the optimization \citep{Tremolet2006} as is needed in standard 4D-Var.  

The optimization window $\tau$ must not be so large that the optimization procedure is computationally infeasible, but large enough to ensure enough points to approximate the conditional variance. Furthermore, it should be large enough so that that inclusion of a new observation at the end of the time window does not influence the initial condition.

\begin{figure}[h]
	\centering\includegraphics[width=1\linewidth]{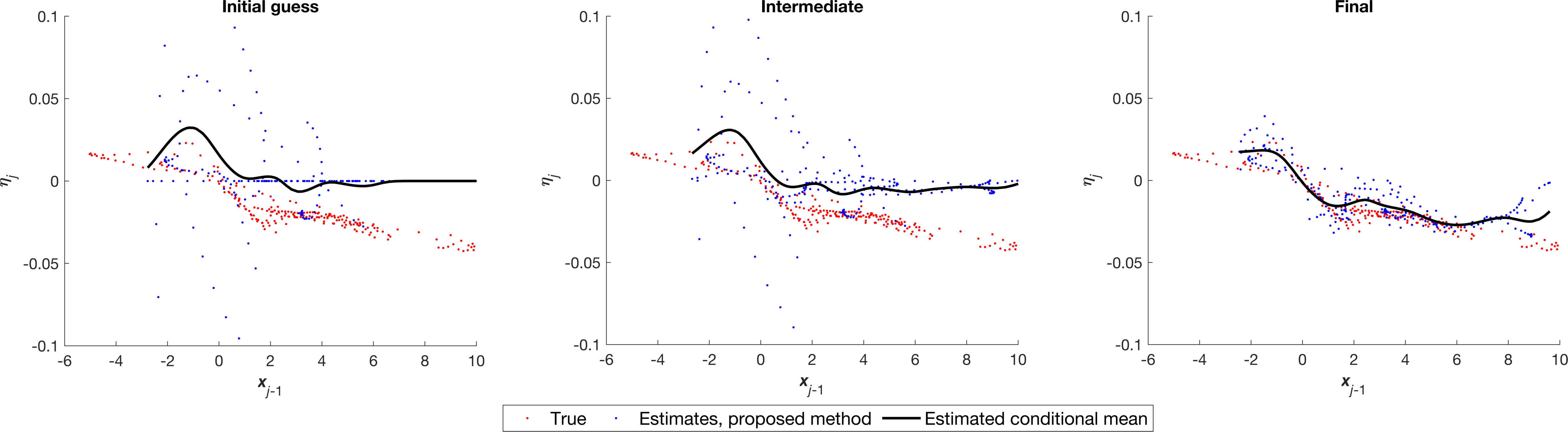}
	\caption{Representative example of the minimization process for a single window.  An iterative minimization algorithm is used starting with an initial guess of zero for unobserved variables. Results are shown at various stages (a - initial guess, b - intermediate and c - final).  The aim is to minimize the deviation of the errors from the nonparametric estimate of the conditional mean, subject to the constraint that the estimated states match the observations.  Notice how the spread of the errors gradually becomes smaller from a) to c).}
	\label{fig:methschem}
\end{figure}

This step of estimating model errors given the observations in the training period is summarized in Algorithm \ref{alg:alg1} and \ref{alg:alg2}.  A pictorial representation of the optimization over a single time window is provided in Figure \ref{fig:methschem}.  It shows the estimated errors at various stages of the iterative minimization process for the numerical experiment considered in Section \ref{S:4}.

\begin{algorithm}
	\caption{Model Error Estimation over training period with sliding window}
	\label{alg:alg1}
	\begin{algorithmic}[1]
		\State {Set: \begin{itemize}
				\item window size, $\tau$
				\item initial state, $\hat{\bm{x}}_0$
				\item total time series length, $T$
				\item initial guess for errors on unobserved variables, ${\bm{\gamma}}^{(0)}_{1:T}$ 
			\end{itemize}}
		\For{$t=1:T-\tau$}
			\begin{align*}
				\bm{\gamma}^{(t)}_{t:t+\tau} & = \text{kernelopt}  \left(\hat{\bm{x}}_{t-1}, {\bm{\gamma}}^{(t-1)}_{t:t+\tau} \right)\\
				\hat{\bm{\eta}}_{t}[k_o] &= \bm{y}_t - \mathbf{H}M(\hat{\bm{x}}_{t-1})  \\
				\hat{\bm{\eta}}_t[k_u] &= {\bm{\gamma}}^{(t)}_t \\
				\hat{\bm{x}}_t &= M(\hat{\bm{x}}_{t-1})  + \hat{\bm{\eta}}_t \\
 {\bm{\gamma}}_{t+1:t+1+\tau}^{(t)} &= \{{\bm{\gamma}}^{(t)}_{t+1:t+\tau}, {\bm{\gamma}}^{(0)}_{t+1+\tau}\}
			\end{align*}
				
	\EndFor	
  \For{$t=T-\tau+1:T$}
	\begin{align*}
		\hat{\bm{\eta}}_{t}[k_o] &= \bm{y}_t - \mathbf{H}M(\hat{\bm{x}}_{t-1})  \\
		\hat{\bm{\eta}}_t[k_u] &= {\bm{\gamma}}^{(t)}_t \\
		\hat{\bm{x}}_t &= M(\hat{\bm{x}}_{t-1})  + \hat{\bm{\eta}}_t
	\end{align*}
  \EndFor
		
	\State \Return { $\hat{\bm{\eta}}_{1:T}$ ; $\hat{\bm{x}}_{0:T-1}$ }
	\end{algorithmic}
\end{algorithm}

\begin{algorithm}

  \caption{kernelopt } 
    \label{alg:alg2}
    \begin{algorithmic}[1]
    \State {Set: \begin{itemize}
				\item state estimate at time $t-1$, $\hat{\bm{x}}_{t-1}$
				\item best guess for unobserved errors for time $t$ to $t+\tau$, $\bm{\gamma}_{t:t+\tau}^{(t-1)}$ 
				\item stopping criterion $J_{stop}$
				\item Kernel function and bandwidth $K_b$ and $b$ respectively
				\item maximum no. of iterations $maxiter$
			\end{itemize}}
	\State {Initialise: \begin{itemize}
	    \item $m = 1$
	    \item $\bm{\gamma}_{t:t+\tau}^{(t)} = \bm{\gamma}_{t:t+\tau}^{(t-1)}$
	\end{itemize}	}
	
	\While{$m < maxiter$}
        \For{$j=t:t+\tau$}
				\begin{align*}
				    \hat{\bm{\eta}}_j[k_o] &= \bm{y}_j - \mathbf{H}M(\hat{\bm{x}}_{j-1}) \\
				    \hat{\bm{\eta}}_j[k_u] &= {\bm{\gamma}}^{(t)}_j \\
				    \hat{\bm{x}}_j &= M(\hat{\bm{x}}_{j-1})  + \hat{\bm{\eta}}_j
				\end{align*}
			
		\EndFor		
		
	\State{Calculate using conditionals $\hat{\bm{z}}_{j,k}$ for all $j,k$ using $\hat{\bm{x}}_{t-1:t+\tau}$:}
        
        \begin{align*}
        \hat{m}(\bm{z}_{j,k}) & = \frac{\sum_{l=1}^K \sum_{i=t}^{t+\tau} K_b(\norm{\hat{\bm{z}}_{i,l} - \hat{\bm{z}}_{j,k}}) \hat{\bm{\eta}}_i[l]}{\sum_{l=1}^K \sum_{i=t}^{t+\tau} K_b(\norm{\hat{\bm{z}}_{i,l} - \hat{\bm{z}}_{j,k}})} \\
	    J(\bm{\gamma}_{t:t+\tau}^{(t)}) & = \sum_{k=1}^K \sum_{j = t}^{t+\tau} \left(\hat{\bm{\eta}}_j[k] - \hat{m}(\hat{\bm{z}}_{j,k}) \right)^2 
		\end{align*}
		
		\If{$J > J_{stop}$}
		\State{Calculate: new guess  
		$\bm{\gamma}_{t:t+\tau}^{(t)}$ as per chosen optimization scheme}
		\State{Set: $m = m+1$}
		\Else 
		\State{Set: $m = maxiter$}
		\EndIf
			
	\EndWhile		

	\State \Return { $\bm{\gamma}_{t:t+\tau}^{(t)}$ }
    \end{algorithmic}
\end{algorithm}

\subsubsection{Step 2. Conditional PDF estimation and sample generation}
The resulting sample of additive error and states estimates $\hat{\bm{\eta}}_{1:T}$ and $\hat{\bm{x}}_{0:T-1}$ from Algorithm \ref{alg:alg1} is now used to derive the conditional probability density e.g. $p(\bm{\eta}_j[k]|\bm{x}_{j-1}[k])$ for a given grid point $k$. Kernel conditional density estimation methods \citep{Hyndman1996, Hall2004} are well suited to such a task, although they are generally data-intensive and suffer from the curse of dimensionality.  However, they are sufficient for the class of problems considered herein where the locality assumption greatly reduces the dimension of the response variable and covariates. We adopt the method of \citet{Hayfield2008} as implemented in the np package in R.  For a set of $N$ data points $\{\bm{x}_i, y_i\}_{i=1:N}$ for covariate $\bm{x} \in \mathbb{R}^d$ and response variable $y \in \mathbb{R}$, a Kernel estimate of the conditional density is constructed as 
\begin{align*}
    \hat{p}(y| \bm{x}) = \frac{\sum_{i=1}^N K_{b_y}(y - y_i) K_{b_x}(\norm{\bm{x} - \bm{x}_i})}{\sum_{i=1}^N K_{b_x}(\norm{\bm{x} - \bm{x}_i})}
\end{align*} 
where $K_b$ is a user specified Kernel function with bandwidth $b$ and $b_x$ and $b_y$ refer to the bandwidths selected for the covariates and response variable, respectively. 

As mentioned earlier, it is also possible to include additional covariates that strongly influence the errors at the current time (e.g. $\bm{\eta}_{j-1}[k]$ to capture serial dependence, as demonstrated in the numerical experiments in Section \ref{S:4}.  The set of covariates is likely to be problem dependent; prior knowledge of the system is required to select them appropriately. 

\subsection{Benchmark Methods}
\label{S:32}
The stochastic parameterization methods for multi-scale systems discussed in Section 1 \citep[][e.g]{Wilks2005, Crommelin2008, kwasniok2012, Arnold2013, Lu2017} require knowledge of the sub-grid scale processes and/or fully observed resolved variables.  These approaches are therefore inapplicable for the problem setting considered here.  In the remainder of this section, two existing data assimilation based methods that are amenable to our problem setting are discussed.  They are also adopted as benchmarks for comparison with the proposed approach.
 
\subsubsection{B1 - Analysis Increment Based Method}
\label{S:321}
Several researchers have investigated the potential of using analysis increments from a data assimilation run to characterize model errors, see for example \citep{Leith1978, Li2009, Mitchell2015}.  We adopt the recently proposed ETKF-TV of \citet{Mitchell2015} as a representative method of such approaches (hereafter referred to as Method B1). Their method consists of estimating a Gaussian model error distribution by calculating the mean and covariance of the analysis-forecast differences over a so-called reanalysis period, via:
  
\begin{linenomath}
\begin{gather}
\label{eq:3l}
\delta\bm{x}_j^a = \frac{1}{n}\sum_{i=1}^{n} \left(\bm{x}_j^{ai} - \bm{x}_j^{fi}\right)\\
\label{eq:3m}
\overline{\bm{b}} = \frac{1}{T} \sum_{j=1}^{T} \delta\bm{x}_j^a \\
\label{eq:3n}
\overline{\mathbf{P}} = \frac{1}{T-1} \sum_{j=1}^{T} \left[\delta\bm{x}_j^a - \overline{\bm{b}}\right] {\left[\delta\bm{x}_j^a - \overline{\bm{b}}\right]}^T 
\end{gather}
\end{linenomath}
where $\bm{x}_j^{ai}$ and $\bm{x}_j^{fi}$ refer to the $i$th ensemble member at time $j$ obtained from the reanalysis assimilation run for the analysis and forecast respectively.  Note (\ref{eq:3l})-(\ref{eq:3n}) are derived assuming the analysis interval length is the same in the reanalysis and experimental run (as is the case in this study).  
This model error distribution is then used to draw model error samples for the the actual data assimilation experiments. Since this estimate of the model error includes analysis errors the authors include a tuning parameters $\alpha$, leading to a model forecast of the form:

\begin{linenomath}
\begin{gather}
\label{eq:3j}
\bm{x}_j^{fi} = M(\bm{x}_{j-1}^{ai}) + \alpha \bm{\eta}_j^i \\
\label{eq:3k}
\bm{\eta}_j^i \sim N(\overline{\bm{b}}, \overline{\mathbf{P}})
\end{gather}
\end{linenomath}

\subsubsection{B2 -Error estimation using Long Window Weak Constraint 4D-Var}
\label{S:322}
As discussed in Section \ref{S:31}, the proposed method for estimation of additive errors relies on ideas from Long window weak-constraint 4D-Var \citep{Tremolet2006} to avoid specification of the background covariance matrix.  However, it differs in the specification of the cost function, as the 4D-Var method provides the least square solution for the model error control variable.  The second benchmark (Method B2) is taken to be the same as the proposed approach, but with Step 1 (see Section \ref{sec:step1}) replaced by Long window weak constraint 4D-Var estimates for the model error.  The probability density estimation (Step 2) remains unchanged. It is worth noting that this is not exactly a "standard" method in its entirety, but is investigated to examine the benefit of the conditional sum of squared deviation minimization aspect of the proposed method.  The long window weak constraint 4D-Var method is discussed below.\\    

In variational data assimilation model errors are accounted for using weak constraint 4D-Var.  In the formulation where the initial state and model errors are considered as control variables, this amounts to minimizing the following cost function over a time window of length $\tau$ \citep{Tremolet2006}:

\begin{linenomath}
\begin{equation} \label{eq:3p}
\begin{split}
J(\bm{x}_0, \bm{\eta}) = \underbrace{\frac{1}{2} \left(\bm{x}_0 - \bm{x}_b \right)^T \mathbf{B}^{-1} \left(\bm{x}_0 - \bm{x}_b \right)}_\textrm{\textit{J\textsubscript{B}}} + \underbrace{\frac{1}{2} \sum_{j=1}^{\tau} \bm{\eta}_j^T \mathbf{Q}_j^{-1}  \bm{\eta}_j}_\textrm{\textit{J\textsubscript{Q}}} \\
+\underbrace{\frac{1}{2} \sum_{j=0}^{\tau} \left(\mathbf{H}\bm{x}_j - \bm{y}_j \right)^T \mathbf{R}^{-1} \left(\mathbf{H}\bm{x}_j - \bm{y}_j \right)}_\textrm{\textit{J\textsubscript{O}}} 
\end{split}
\end{equation}
\end{linenomath}
where $\bm{x}_b$ is the background estimate of the initial state $\bm{x}_0$ ; $\mathbf{B}$ is the background error covariance matrix associated with $\bm{x}_b$; $\mathbf{Q}$ is the model error 
covariance matrix; and $\bm{x}_j=M(\bm{x}_{j-1})+ \bm{\eta}_j$ for $j  \in \{1,…\tau\}$.  The assimilation cycle is repeated by then considering the next assimilation window $\{\tau+1,…2\tau\}$.  However, in the long window approach, minimization is performed by shifting the interval by one observation interval rather than the full assimilation window of length $\tau$.  This allows one to neglect the background term $J_B$ from the cost function after a suitable warm up period.  The estimate $\bm{x}_b$ would have already converged due to the many iterations of the minimisation algorithm from the overlapping windows, meaning its uncertainty is negligible in comparison to the other terms.  The window length should also be chosen to be sufficiently long, such that the inclusion of a new observation at the end of the time window does not affect the initial state (this is relevant for the proposed approach also).\\

In summary, the B2 method is defined as being the same as the proposed approach, but with the cost function in (\ref{eq:new14}) replaced by minimization of the following cost function for any given time $t$:
\begin{linenomath}
\begin{equation}
\label{eq:3q}
J(\bm{\eta}_{t:t+\tau}, \hat{\bm{x}}_{t-1}) =  \underbrace{\frac{1}{2} \sum_{j=t}^{t+\tau} \bm{\eta}_j^T \mathbf{Q}_j^{-1}  \bm{\eta}_j}_\textrm{\textit{J\textsubscript{Q}}} \\
+\underbrace{\frac{1}{2} \sum_{j=t}^{t+\tau} \left(\mathbf{H}\bm{x}_j - \bm{y}_j \right)^T \mathbf{R}^{-1} \left(\mathbf{H}\bm{x}_j - \bm{y}_j \right)}_\textrm{\textit{J\textsubscript{O}}} 
\end{equation}
\end{linenomath} 
where $\bm{x}_j=M(\bm{x}_{j-1})+ \bm{\eta}_j$ for $j  \in \{t,…t+\tau\}$ and the initial condition is given by $\hat{\bm{x}}_{t-1}$.
The estimated model error distribution is derived in the same way as for the proposed method, as detailed below.

The Levenberg-Marquardt algorithm is again used as the minimizer, for the sake of comparison with the proposed approach.  Notice that unlike the proposed approach, the errors in the entire state vector (not just unobserved states) must be optimized.    

\section{Numerical Experiments}
\label{S:4}

\subsection{Multi-Scale Lorenz 96}
\label{S:41}
Here we investigate the efficacy of the proposed method and benchmarks discussed in Section \ref{S:3} through synthetic experiments using the multi-scale Lorenz 96 model.  This system has been used extensively as a toy model of the atmosphere to test new algorithms and to study model errors due to unresolved sub-grid processes. It consists of a coupled system of equations representing the evolution of an atmospheric quantity discretized over a latitude circle at different scales:
\begin{linenomath}
\begin{gather}
\label{eq:4a}
\frac{dX_k}{dt} = -X_{k-1}(X_{k-2} - X_{k+1}) - X_{k} + F + U_k 
;  \: k \in \{1, ..., N_x\}\\
\label{eq:4b}
\xi \frac{dZ_{l,k}}{dt} = -Z_{l+1,k}(Z_{l+2,k} - Z_{l-1,k}) - Z_{l,k} + h_zX_k ; \quad l \in \{1, ..., N_z\} 
\end{gather}
\end{linenomath}
The ${\{X_k\} }_{k=1}^{N_x}$ 
variables represent quantities evolving on a coarse spatial scale with low-frequency large amplitude fluctuations, where the subscript $k$ refers to the $k$th grid point on the latitude circle.  Each $X_k$ variable is coupled to $N_z$ small-scale variables $Z_{l,k}$ that are characterized by a high frequency and relatively small amplitude 
evolution.  The variables are driven by a quadratic term that models advection, a linear damping, constant forcing $(F)$ and coupling terms that link the two scales.   The system is subject to periodic boundary conditions, so that $X_k= X_{k+k}, Z_{l,k}= Z_{l,k+N_x} $ and $Z_{l+N_z,k}= Z_{l,k+1}$.  The effect of the unresolved fast variables on the slow variables is denoted by the so-called sub-grid tendency $U_k$:
\begin{linenomath}
\begin{equation}
\label{eq:4c}
U_k = \frac{h_x}{N_z} \sum_{l=1}^{N_z} Z_{l,k}
\end{equation}
\end{linenomath}
We use the formulation of the Lorenz 96 (\ref{eq:4a})-(\ref{eq:4b}) as provided in \citep{Fatkullin2004} which makes the time-scale separation between the slow and fast variables (measured by $\xi$) explicit.  Note that this formulation is equivalent to the system originally proposed by Lorenz with the following 
parameter conversions: 
$\xi=  \frac{1}{c}$ where $c$ = time scale ratio; $h_x= \frac{-hcN_z} {b^2}$  where $b$ = spatial scale ratio and $h$ is the coupling constant; and $h_z=h$.\\

\begin{figure}[h]
	\centering\includegraphics[width=1\linewidth]{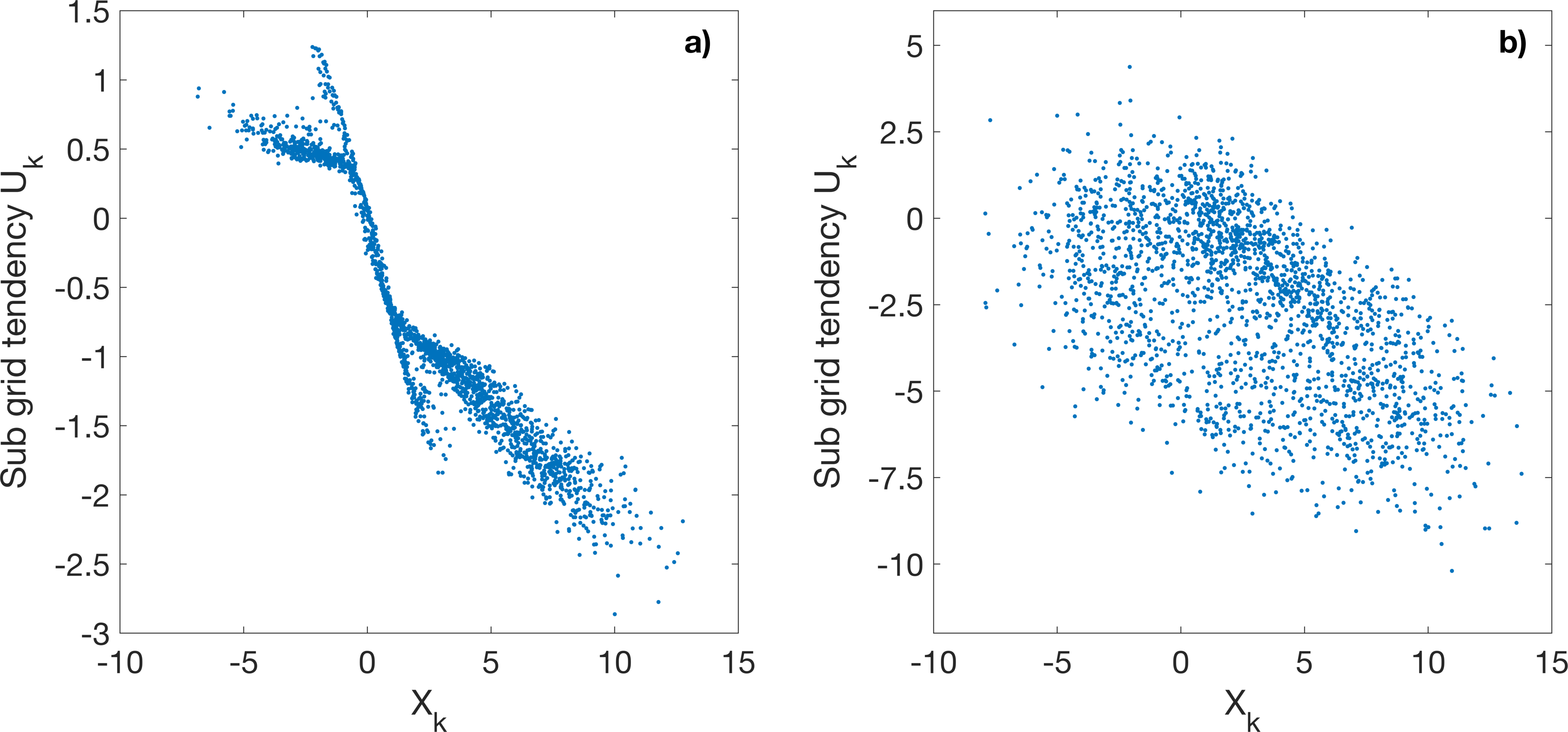}   
	\caption{Sub-grid tendencies for the two different regimes of the multi-scale Lorenz 96 system considered in this study: a) Case 1 - large time scale separation; b) Case 2 - small time scale separation.  For both cases, points are sampled at an interval of 0.3 MTU} 
	\label{fig:subgridtend}
\end{figure}

The behaviour of the system can vary considerably depending on the values assigned to the parameters in (\ref{eq:4a})-(\ref{eq:4b}).  We consider 
two dynamical regimes to study the robustness of the proposed approach to different model error structures, summarised in Table \ref{table:expsetup}.  We first consider a case with large time scale separation $(\xi \approx 0.008)$ studied by \citet{Fatkullin2004}. The sub-grid tendency has a complex non-linear dependence on the resolved variable, making it of interest to this study.  However, such large time scale separations are not representative of the real atmosphere.  We therefore consider a second case that has a much smaller time scale separation $(\xi \approx 0.7)$ and stronger coupling $(h_x=-2)$ than cases considered in previous studies (where $\xi$ is typically 0.4 or 0.5 and $h_x=1$) \citep[e.g.][]{Crommelin2008, Arnold2013, Lu2017}.  Smaller values of $\xi$  (i.e. larger time scale separation) are generally considered more difficult to parameterize, and the larger magnitude of the coupling term amplifies the effect of model errors.  In both case studies, the dynamics are chaotic and give rise to complex non-Gaussian conditional error densities, as shown by the variation of $U_k$ with $X_k$ in Figure \ref{fig:subgridtend}. 

\begin{table}[h]
	\centering
	\begin{tabular}{>{\centering\arraybackslash}m{3cm} | >{\centering\arraybackslash}m{3cm} | >{\centering\arraybackslash}m{3cm} |  >{\centering\arraybackslash}m{3cm}}
		\hline
		\multirow{6}{3cm}{Lorenz 96 parameters} & \textbf{Parameter} & \textbf{Case Study 1} & \textbf{Case Study 2} \\
		\hline
		& $\xi$ & $\frac{1}{128} \approx 0.008$ & 0.7 \\
		&$h_x$ & -0.8 & -2 \\
		&$h_z$ & 1 & 1 \\
		&$N_z$ & 128 & 20 \\
		&$N_x$ & 9 & 9 \\
		&$F$ & 10 & 14\\
		\hline
		\multirow{2}{3cm}{Observation density} & Observation frequency (MTU) & 0.02 & 0.04 \\
		&Location of Observed $X_k$ & $[3,4,8,9]$ & $[1,2,5,6]$\\
		\hline
	\end{tabular}
	\caption{Multi-scale Lorenz 96 parameters for the 2 different case studies.  Note that 1 $ \text{MTU} = \frac{1}{\Delta t} \enskip \text{time steps}$ and the model equations are discretized with $\Delta t = 8 \times 10^{-4}$.}
	\label{table:expsetup}
\end{table}

\subsection{Experimental Setup}
\label{S:42}
The available forecast model is the single scale Lorenz 96 model (\ref{eq:4d}), where the forcing term is known perfectly but knowledge of the sub-grid processes $Z_{l,k}$ is unavailable:  
\begin{linenomath}
	\begin{equation}
	\label{eq:4d}
	\frac{dX_k}{dt} = -X_{k-1}(X_{k-2} - X_{k+1}) - X_{k} + F ;  \quad k \in \{1, ..., N_x\} \\
	\end{equation}
\end{linenomath}

Our aim is to first characterize the uncertainty in model simulations due to missing physics i.e. the subgrid term in (\ref{eq:4c}), where the resolved variables are partially observed.  Then we study the effects of uncertainty characterization on forecasts and assimilation.   

\subsubsection{Training Period}
\label{S:421}
 
A truth run for the training period was first generated by numerically integrating the full multi-scale system (\ref{eq:4a})-(\ref{eq:4b}) using a fourth-order Runge-kutta scheme with time step $\Delta t=0.0008$.  Similar to \cite{Arnold2013}, we use MTU to denote model time units, where $1 MTU = \frac{1}{\Delta t}$. 

Partial observations of the resolved slow variables were then developed by perturbing the true values with zero mean, temporally and spatially uncorrelated Gaussian noise:
\begin{linenomath}
\begin{equation}
\label{eq:4e}
\begin{gathered}
\bm{y}_j = \mathbf{H}\bm{x}_j + \bm{\varepsilon}_j \\
\bm{\varepsilon}_j \sim N(0, \mathbf{R})
\end{gathered}
\end{equation}
\end{linenomath}
where $\mathbf{H}$ is a non-square matrix of 1s and 0s with $N_y < N_x$, $\bm{x}_j$ is the true state at time $j$ where $\bm{x}_j [k]$  is equivalent to the time discretised value of $X_k$ in (\ref{eq:4a}), and $\mathbf{R} = 10^{-7} \mathbf{I}_{N_y}$ where $\mathbf{I}_{N_y}$ is the identity matrix of size $N_y$.  $\mathbf{R}$ was chosen such that measurement errors are negligible in comparison to model errors. 

The resolved variables are partially observed in space in all experiments (approx. 50\% observed, see Table \ref{table:expsetup}), and observations are available at 0.02 and 0.04 MTU for Case Studies 1 and 2 respectively.  Based on the work of \citet{Lorenz2006}, this corresponds to an observation interval of 2.5 and 5 hrs respectively (1 MTU is approximately equivalent to 5 days).  These interval lengths where chosen to reflect a realistic observation network whilst also maintaining complex non-Gaussian error structures. 

The proposed approach was then applied to the training data.  The forecast model (\ref{eq:4d}) was integrated using a time step of $\Delta t=8 \times 10^{-4}$.  We estimate the following probability densities:  
\begin{center}
	Case Study 1: $p(\bm{\eta}_j[k]|\bm{x}_{j-1}[k])$ \\
	Case Study 2: $p(\bm{\eta}_j[k]|\bm{x}_{j-1}[k], \bm{x}_{j-1}[k-1], \bm{\eta}_{j-1}[k])$ 
\end{center}  
Note the inclusion of the past value of the model error in Case Study 2, which is related to the presence of time correlations in errors in the Lorenz 96 system (as identified by e.g. \citet{Arnold2013}). 
In a real system, such choices would be informed by expert knowledge of the error processes. 

Window lengths equal to $\tau=25$ and 50 observation intervals
were selected for Case Study 1 and 2, respectively.  This was sufficiently long to capture a range of dynamical states and also longer than the system decorrelation time, so that the sliding window approach can be utilized to ignore the background term in the cost function (as discussed in Section \ref{S:322}).  To ensure temporal independence the data for the nonparametric conditional density estimation was generated by sampling the estimated error and states at an interval of 0.3 MTU, where autocorrelation is approximately zero.  A Gaussian Kernel function was adopted throughout using the data-driven bandwidth estimation procedure as detailed in \cite{Hayfield2008}.  To avoid issues related to bandwidth specification and data sparsity in high dimensions, outlier points in the covariate space were removed from the data used for density estimation in Case Study 2.

This training data was also used in the benchmark methods.  For method B2 we used the same window length and density estimation algorithm as for the proposed approach.  The process error covariance matrix $\mathbf{Q}$ was estimated by calculating the sample covariance of the true errors over the training period. For the B1 method the inflation parameter was tuned based on the analysis RMSE, whilst the correction factor $\alpha$ was selected by evaluating the spread vs RMSE relationships, as it has a greater impact on ensemble spread than accuracy.  Localization was not required due to the large ensemble size $(n=1000)$ relative to the state dimension.  

\subsubsection{Assimilation Period}
\label{S:422}
The estimated errors and analysis increments were then explored in assimilation experiments using the ETKF.  The forecast model in the assimilation experiment was also the single scale Lorenz 96 (\ref{eq:4d}); spatio-temporal observation frequency was the same and observations were generated also using (\ref{eq:4e}).  Assimilation was undertaken for 30 independent runs of length 100 observation intervals with independent initial conditions.  Truth runs were first generated using the same approach as for the training period.  The initial conditions were generated by selecting 30 values on the attractor at intervals of 12500 time steps, which is sufficient to ensure that the autocorrelation in the resolved variables is close to zero (also adopted by \citet{Arnold2013}).  Perfect initial conditions were adopted in all experiments as the focus is on the effects of model error.  Similarly, a large ensemble size $(n=1000)$ was adopted to minimize the effects of sampling error and to avoid the use of localization methods.  

\subsection{Results and Discussion}
\label{S:43}

\subsubsection{Model Error Estimation}
\label{S:431}
\begin{figure}[h]
	\centering\includegraphics[width=1\linewidth]{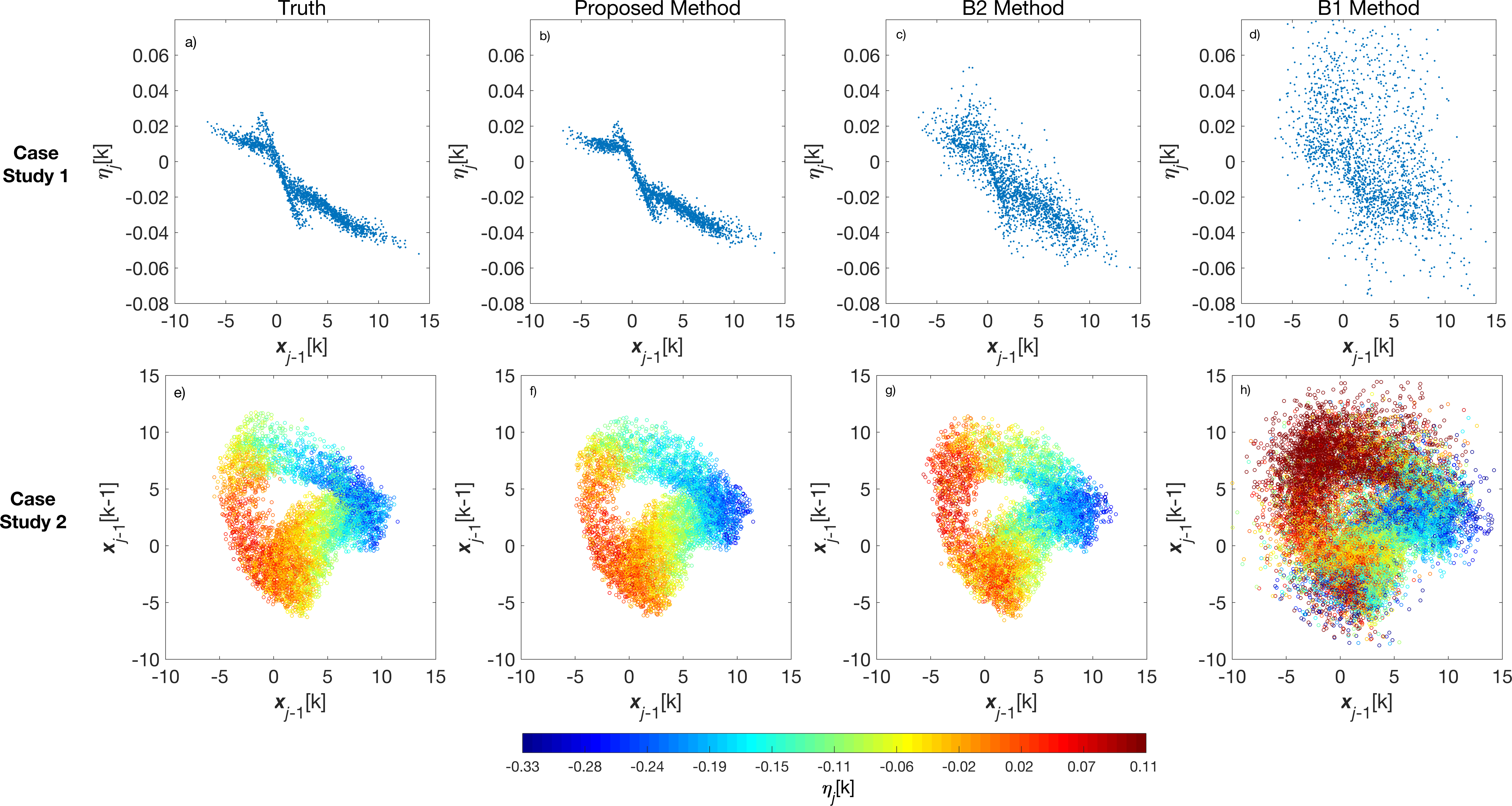}   
	\caption{Sub-grid tendencies for the two different regimes of the multi-scale Lorenz 96 system considered in this study: a) to d) Case 1 - large time scale separation; b) to h) Case 2 - small time scale separation.  For both case studies, points are sampled at an interval of 0.3 MTU} 
	\label{fig:estimatederrs}
\end{figure}

In both case studies, the proposed approach recovers the true error estimates from partial observations more accurately than the benchmark methods.  This is demonstrated qualitatively in Figure \ref{fig:estimatederrs}, which shows the sample set of additive errors $\bm{\eta}_j[k]$ against the spatial covariates (resolved variable at the previous observation time, $\bm{x}_{j-1}[k]$  for Case Study 1; and $\bm{x}_{j-1}[k-1]$  and $\bm{x}_{j-1}[k]$  for Case Study 2).  In Case Study 1, the B2 method manages to at least partially recover the non-linear relationship between $\bm{\eta}_j[k]$ and $\bm{x}_{j-1}[k]$, but is less precise than estimates from the proposed method (compare Figure \ref{fig:estimatederrs}b \& c).  In Case Study 2, it more closely reflects the true error structure, although an overestimation and underestimation of error values is apparent in key regions of the covariate space (compare Figure \ref{fig:estimatederrs}e, f \& g).  Method B1 produces poor quality error estimates in both case studies; errors are grossly overestimated and the dependence structure between the errors and covariates is poorly represented.

\begin{figure}[H]
	\centering\includegraphics[width=1\linewidth]{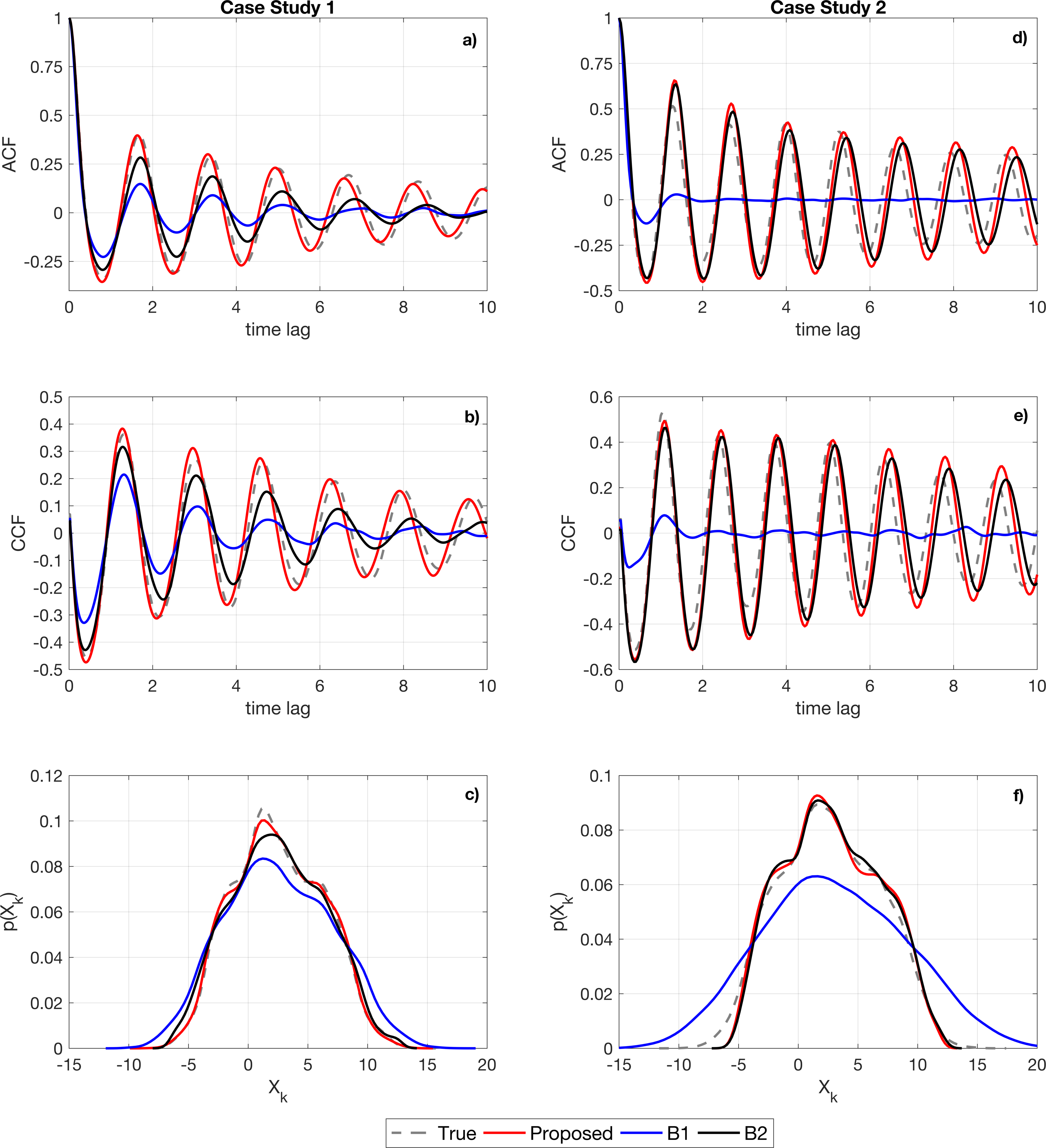}   
	\caption{ACF, CCF and marginal density of a resolved variable for both case studies using different parameterization approaches.} 
	\label{fig:ACFCCF}
\end{figure}

The model error estimation techniques considered here can also be considered as stochastic parameterizations of the sub-grid dynamics. The ability of the methods to replicate key characteristics of the full 2-scale Lorenz 96 model when used in this manner is also assessed. For each case the single-layer Lorenz 96 system is run for $10^5$ time steps with a $\Delta t=8 \times 10^{-4}$, adding draws from the model error pdfs at the observation intervals used to construct these pdfs, i.e. 0.02 and 0.04 MTU for Case Study 1 and 2, respectively. We calculate the autocorrelation function (ACF) of $X_k$, the cross-correlation function (CCF) between $X_k$ and $X_{k+1}$, and the marginal probability density of $X_k$ (see Figure \ref{fig:ACFCCF}).  The correlation functions approximate the dynamical transitions of the slow variables whilst the marginal probability density approximates the invariant measure.  Again, Method B1 performs poorly in all aspects, particularly in Case Study 2 where temporal correlations are not reproduced, meaning that the dynamical transitions are poorly represented.  The results are similar to those from using inflation and localization only, in case studies with a similar time scale separation \citep[e.g][]{Lu2017}. Improvements of the proposed method over Method B2 are more distinct in Case Study 1 than in Case Study 2, consistent with the greater similarity in error estimates in this case study (see Figure \ref{fig:estimatederrs}). The Proposed Method reproduces all three features relatively accurately in both case studies, and even compares favourably with other methods that rely on data of the sub-grid processes (c.f. Figures 5-7 in \citet{Crommelin2008} and Figure 1 in \citet{Lu2017}).  .  
\\

\begin{figure}[h]
	\centering\includegraphics[width=0.6\linewidth]{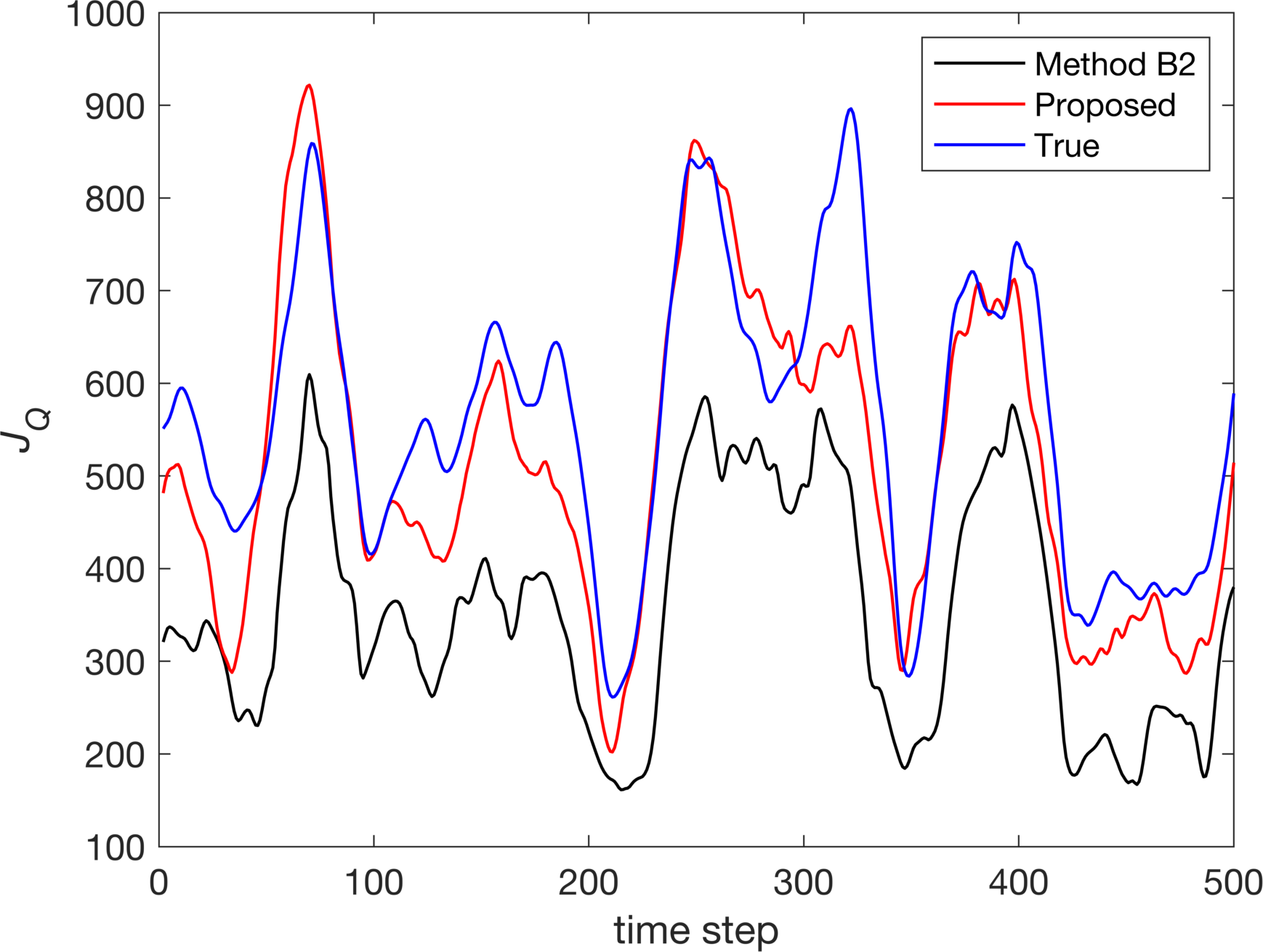}   
	\caption{Snapshot of $J_Q$ values (see (\ref{eq:3q})) for method B2, proposed and the true data for Case Study 2.} 
	\label{fig:Jqvalues}
\end{figure}

The superior performance of the proposed method is attributed to two aspects 1) the formulation of the cost function which aims to minimize the conditional sum of squared deviations of the estimated errors; and 2) optimization of errors over a time window (as is performed in traditional 4D-Var and smoothing methods).  Firstly, minimizing the conditional sum of squared deviations of the errors allows one to estimate more complex state dependent error structures, as opposed to the 4D-Var type approach in Method B2 where dependence information is not taken into account.  The error estimates from Method B2 give $J_Q$ terms (see (\ref{eq:3q})) that are most often lower than $J_Q$ of the true data (see Figure \ref{fig:Jqvalues}), meaning that estimates are obtained by minimizing an inappropriate cost function for this setting.   Furthermore, the proposed approach has the added benefit of avoiding the specification of a model noise covariance matrix, which is needed in Method B2.  

Secondly, optimization over a time window allows one to more effectively constrain the range of possible errors in the partially observed setting, particularly when errors are time correlated.  This partly explains the poor performance of Method B1, which is based on increments from a filter).  Furthermore, the error estimates from Method B1 are heavily influenced by the quality of the assimilation algorithm (ETKF with inflation).  Poorly specified prior uncertainty in the unobserved variables from the inflation procedure can lead to large incremental updates in observed variables at future times.  This ultimately corrupts the error estimates, as the increments are now dominated by initial condition errors in the unobserved variables.  

\subsubsection{Model Error Estimation with Non-Negligble Observation Error}
In the aforementioned experiment, negligible observation errors ($\mathbf{R} = 10^{-7} \mathbf{I}_{N_y}$) were considered, consistent with assumption 3.  This assumption is clearly a limitation for real world applications, and future work will examine how this assumption can be relaxed.  As a first step in this direction, we examined the robustness of the procedure by repeating the error estimation procedure described in Section \ref{S:31} for Case Study 1, but with larger observation error $\mathbf{R} = 10^{-4} \mathbf{I}_{N_y}$.  With this choice, $\frac{\mathbf{Q}_{ii}}{\mathbf{R}_{ii}} \approx 3$, that is the model error variance is approximately 3 times observation error variance.   

\begin{figure}[H]
	\centering\includegraphics[width=1\linewidth]{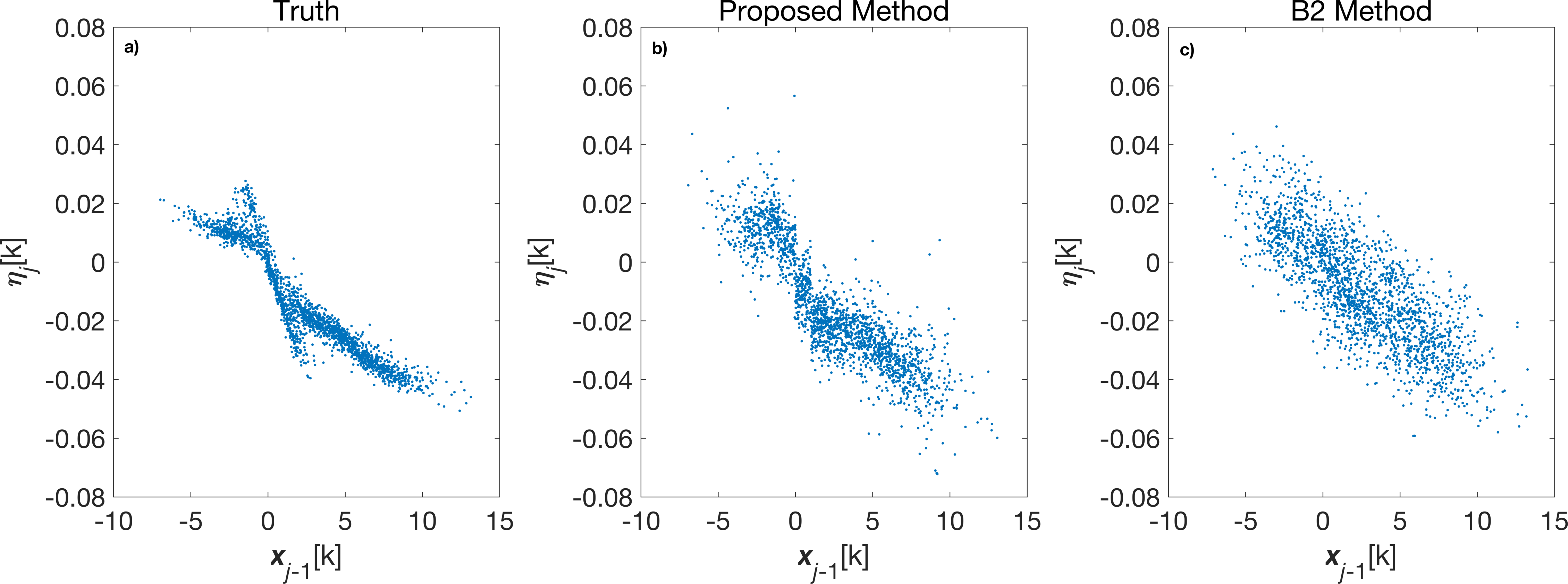}   
	\caption{Sub-grid tendencies for Case Study 1 with increased observation error.} 
	\label{fig:withobserror}
\end{figure}

Figure \ref{fig:withobserror} shows that although error estimates from the Proposed method are not as precise as in the previous experiment, they still capture the underlying error structure more effectively than benchmark B2.  

\subsubsection{Forecast Skill}
\label{S:432}
\begin{figure}[H]
	\centering\includegraphics[width=1\linewidth]{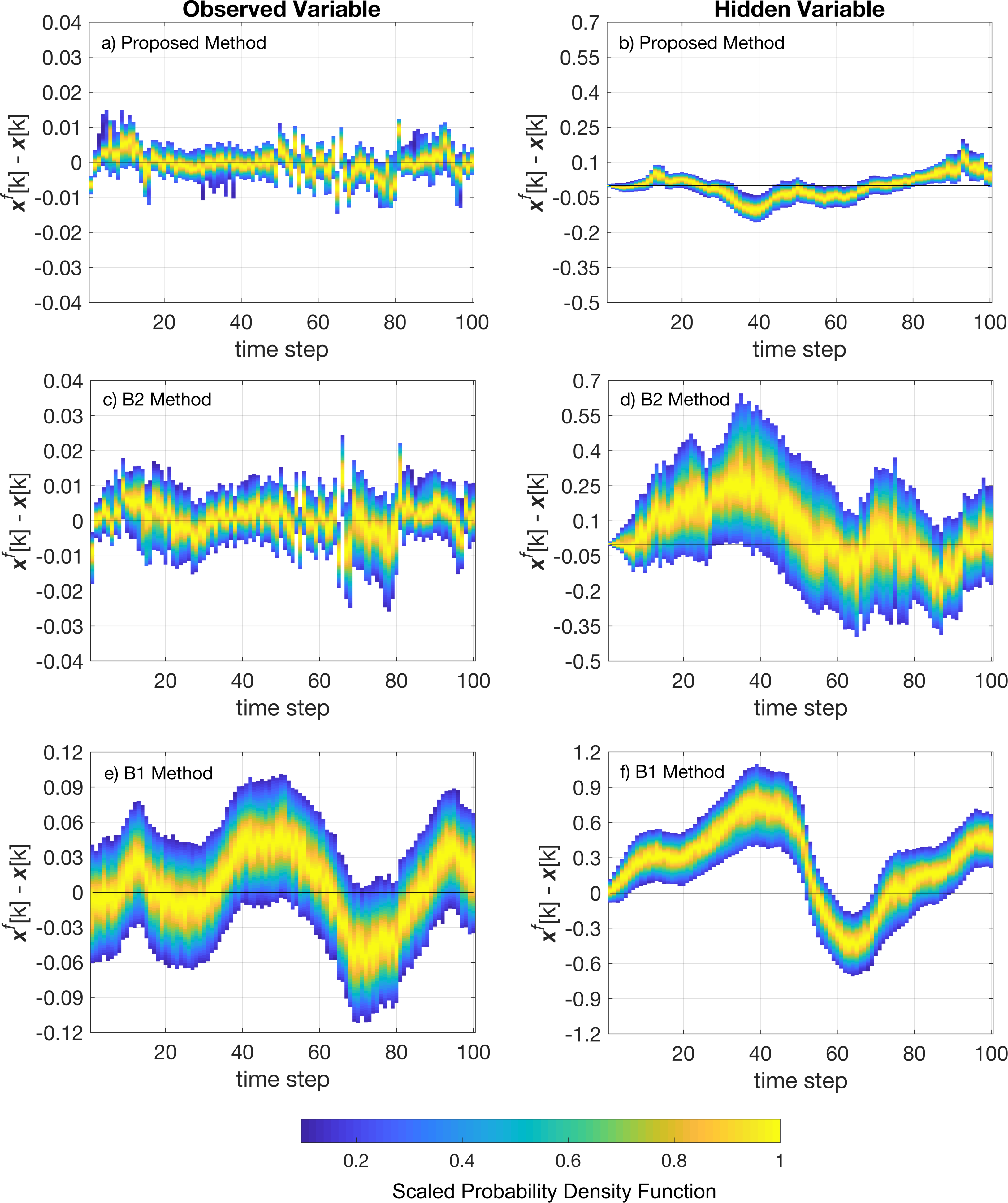}   
	\caption{Representative forecast probability densities of anomalies $(x_t^f [k]- x_t[k])$ for an observed variable (left column) and unobserved variable (right column) for the different methods for Case Study 1.  Forecasts are one-step-ahead (in this case, 0.02 MTU). Each pdf is normalised by its maximum value for the sake of comparison.} 
	\label{fig:fordensCS1}
\end{figure}

\begin{figure}[H]
	\centering\includegraphics[width=1\linewidth]{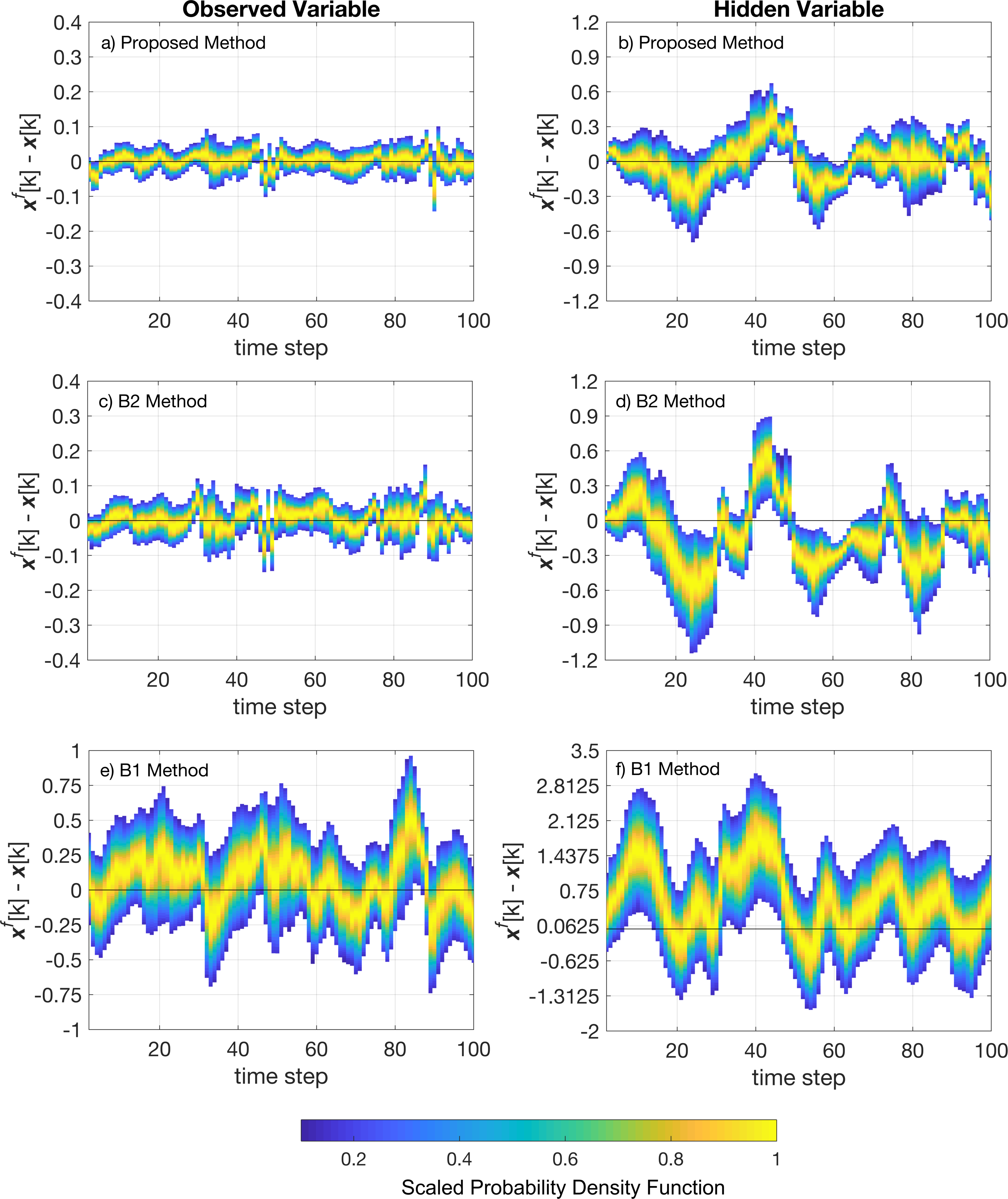}  
	\caption{Representative forecast probability densities of anomalies $(x_t^f [k]- x_t [k])$ for an observed variable (left column) and unobserved variable (right column) for the different methods for Case Study 2.  Forecasts are one-step-ahead (in this case, 0.04 MTU).  Each pdf is normalised by its maximum value for the sake of comparison. } 
	\label{fig:fordensCS2}
\end{figure}

The superior error estimates from the proposed approach leads to improved forecasts compared to the benchmark methods.  Representative results of one-step-ahead forecasts for both case studies are provided in Figure \ref{fig:fordensCS1} and Figure \ref{fig:fordensCS2}.  They show the forecast probability density function (pdf) of the ensemble anomalies (forecast - truth) for both an observed (left column) and unobserved (right column) variable for a single assimilation run of 100 cycles.  In both case studies, relatively large systematic errors can be seen when using Method B1 compared to the other approaches, which is unsurprising given the results in Figure \ref{fig:estimatederrs}.  One step ahead forecasts of the observed variables are relatively similar between the proposed method and the B2 method, although the forecast variance is considerably lower, particularly in Case Study 1.  This is a direct consequence of the more precise additive model error estimates obtained from the proposed approach.  

\begin{figure}[H]
	\centering\includegraphics[width=1\linewidth]{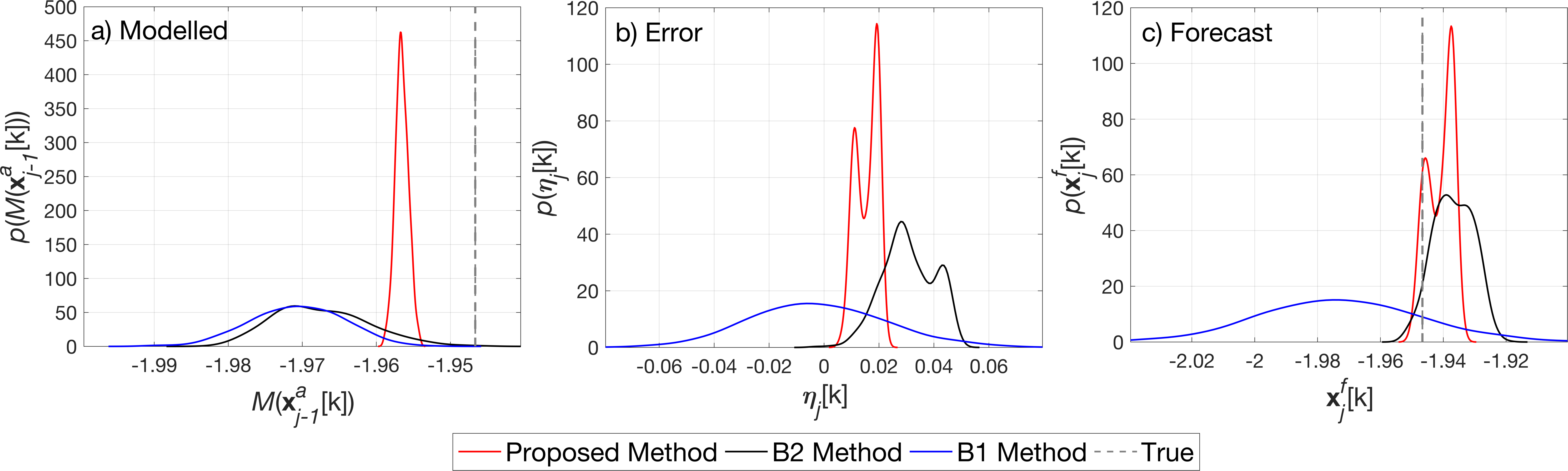}   
	\caption{Example showing benefit of accounting for bimodal transition in an observed variable in Case Study 1 (shown for $t = 82$ in Figure \ref{fig:fordensCS1})} 
	\label{fig:bimodal}
\end{figure}

As an example, resolving bimodality of the transition errors allows one to generate more accurate analyses (hence initial conditions for subsequent time steps) and forecasts, even in an Ensemble Kalman Filter setting.  This is demonstrated in Figure \ref{fig:bimodal} where initial conditions and forecasts in Method B2 have greater variance than in the proposed method, as it is unable to precisely resolve the two modes of the error density.  Differences between forecasts from the proposed and B2 method are much more pronounced for the hidden variables, where both bias and variance are much lower when using the proposed method in both case studies.  The conditional sum of squared deviations minimization procedure allows for a more accurate representation of the spatial dependence structure.  This means that information from observed variables is more effectively transferred to unobserved variables during the assimilation, thereby contributing to the improved forecasts seen for the Proposed Method compared to Method B2, through better initial conditions.  Forecast skill is assessed quantitatively in the remainder of this section.\\
 
A range of forecast metrics were considered to quantify forecast properties including reliability, resolution, accuracy and consistency.  Reliability and resolution were quantified using the Continuous Ranked Probability Score (CRPS) and the (negative) Logarithmic Score (LS) given in (\ref{eq:4j})-(\ref{eq:4l}):

\begin{linenomath}
	\begin{gather}
	\label{eq:4j}
	CRPS_j = \int_{-\infty}^{\infty}  {\left(F_j^f(y) - F_j^o(y)\right)}^2 dy \\
	\label{eq:4k}
	F_j^o(y) = \begin{cases}
	0 \qquad y < y_j \\
	1 \qquad y \geq y_j 
	\end{cases}\\
	\label{eq:4l}
	LS_j = -\ln{\left(p_j^f(y = y_j)\right)}
	\end{gather}
\end{linenomath}
where $F_j^f (y)$ is the empirical cumulative distribution function of the forecast of variable $y$ at time $j$; $F_j^o (y)$ is the cumulative distribution function of the observations of $y$ at time $j$; and $p_j^f (y=y_j)$ indicates the value of the forecast probability density function, evaluated at the observation value.  For cases where only a single observation of $y$ is available at each time, the Heaviside step function is used to characterize the cumulative distribution function of the observation (see (\ref{eq:4k})).\\

The CRPS is routinely adopted in forecasting studies, although it can be a poor statistic for complex forecast probability densities (see for example \citet{Smith2015} who showed that the CRPS can give misleadingly good scores to outcomes that fall in between two modes of a bimodal forecast density).  Hence, the LS is also calculated, although it has the drawback of heavily penalising forecasts in which the outcome falls outside the forecast range.  Accuracy is measured by the Root Mean Squared Error (RMSE), which is evaluated on the ensemble mean.  

Statistical consistency is characterized using RMS Error vs. RMS Spread diagnostic plots, which has been adopted in similar studies \citep[see e.g.][]{Arnold2013}. Ensemble forecasts are considered statistically consistent if the expected ensemble variance equals the expected squared ensemble mean error (assuming unbiasedness and a large enough ensemble size). We separate forecasts into 10 equally populated bins according to their forecast variance, and the mean square spread and mean square error are calculated for each bin prior to taking the square root.  

We used the forecast skill score $(FSS)$ to quantify the relative improvement of the proposed approach over the benchmark methods, defined as:  
\begin{linenomath}
\begin{equation}
\label{eq:4m}
FSS = \frac{Score_{Pr} - Score_{Be}}{Score_{Pe} - Score_{Be}}
\end{equation}
\end{linenomath}
where $Score_{Pr}$ indicates the forecast score of the proposed method; $Score_{Be}$ indicates the forecast score of the reference method (i.e. Method B1 or B2); and $Score_{Pe}$ indicates the score associated to a perfect forecast (e.g. a perfect forecast has FSS = 1).  A skill score of 0.5 means that the proposed approach provides a 50\% improvement over the benchmark, whilst a negative score indicates a degradation in performance. \\

\begin{figure}[H]
	\centering\includegraphics[width=1\linewidth]{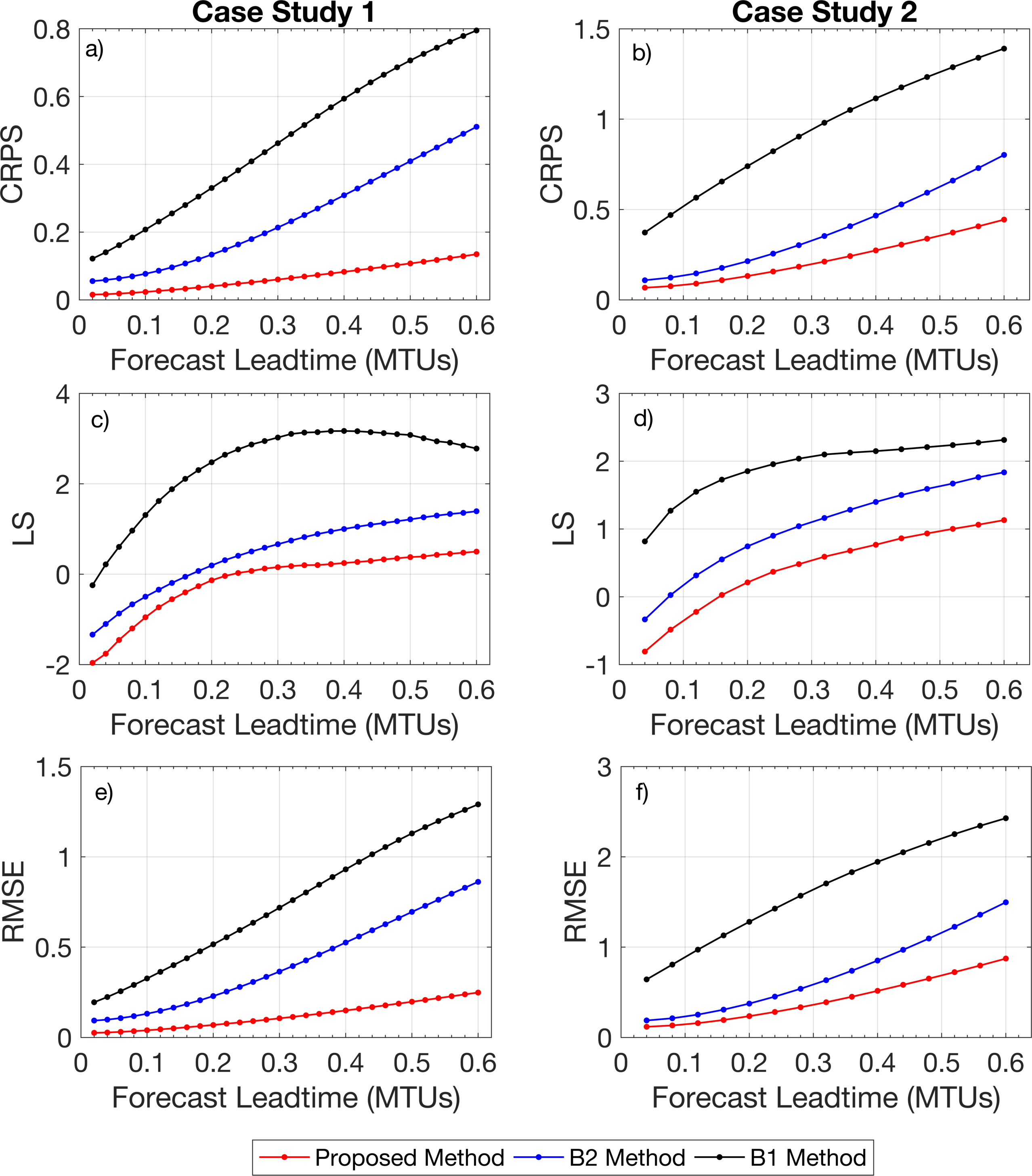}   
	\caption{Forecast scores against lead time for both case studies.  Scores are presented as averages across space (i.e. over all $k$ variables) and across all simulations.  Lower values indicate better performance.}
	\label{fig:rawscores}
\end{figure}

\begin{figure}[H]
	\centering\includegraphics[width=1\linewidth]{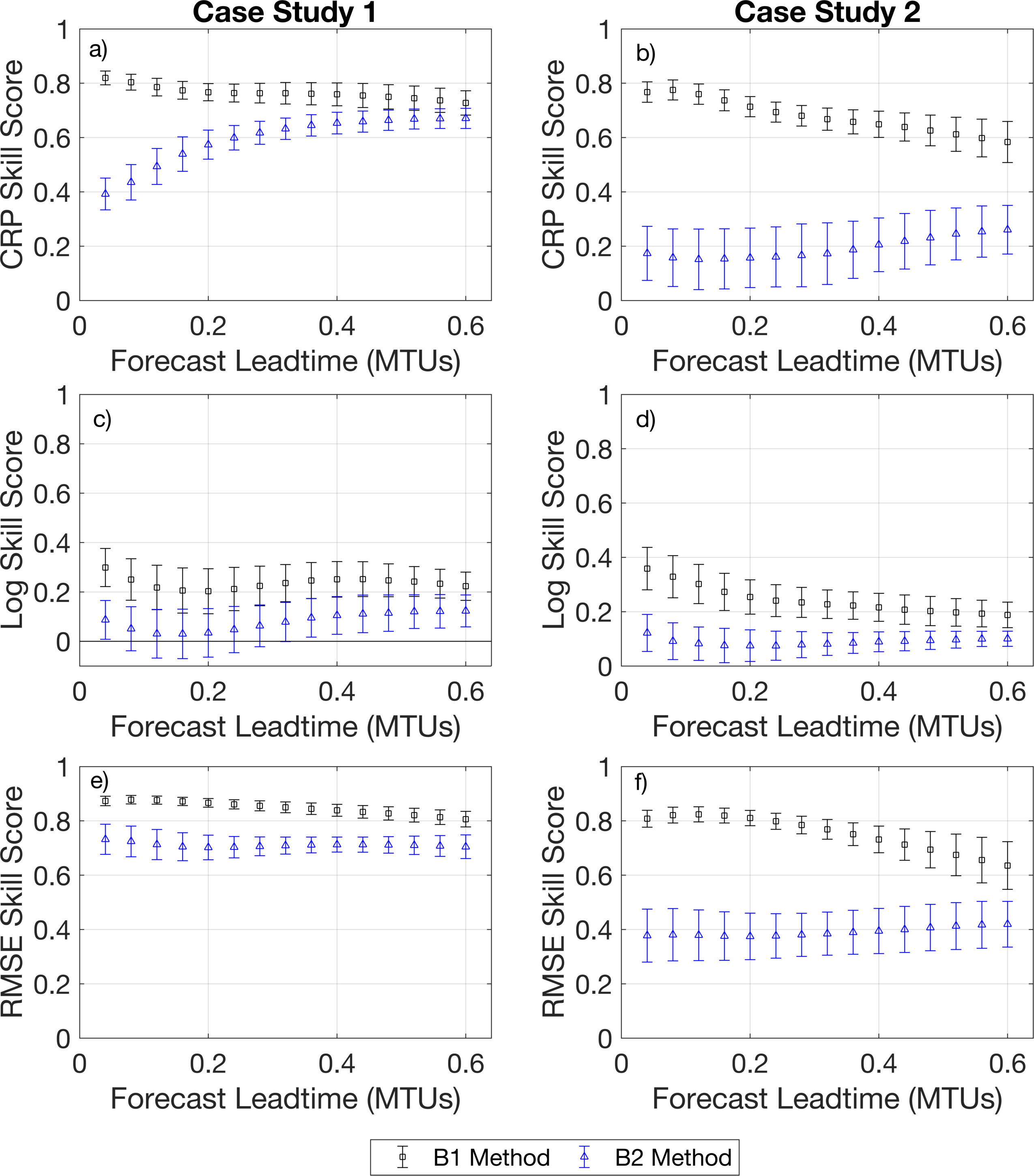}  
	\caption{Forecast skill scores against lead time for both case studies.  Skill scores are first averaged across space (i.e. over all $k$ variables) and time within each independent simulation.  The average of all such values over the 30 independent simulations is shown in the plot (square and triangle markers), as well as the standard deviation.  More positive skill scores indicate greater relative improvement of the Proposed method compared to the benchmark method.} 
	\label{fig:skillscores}
\end{figure}

\begin{figure}[H]
	\centering\includegraphics[width=1\linewidth]{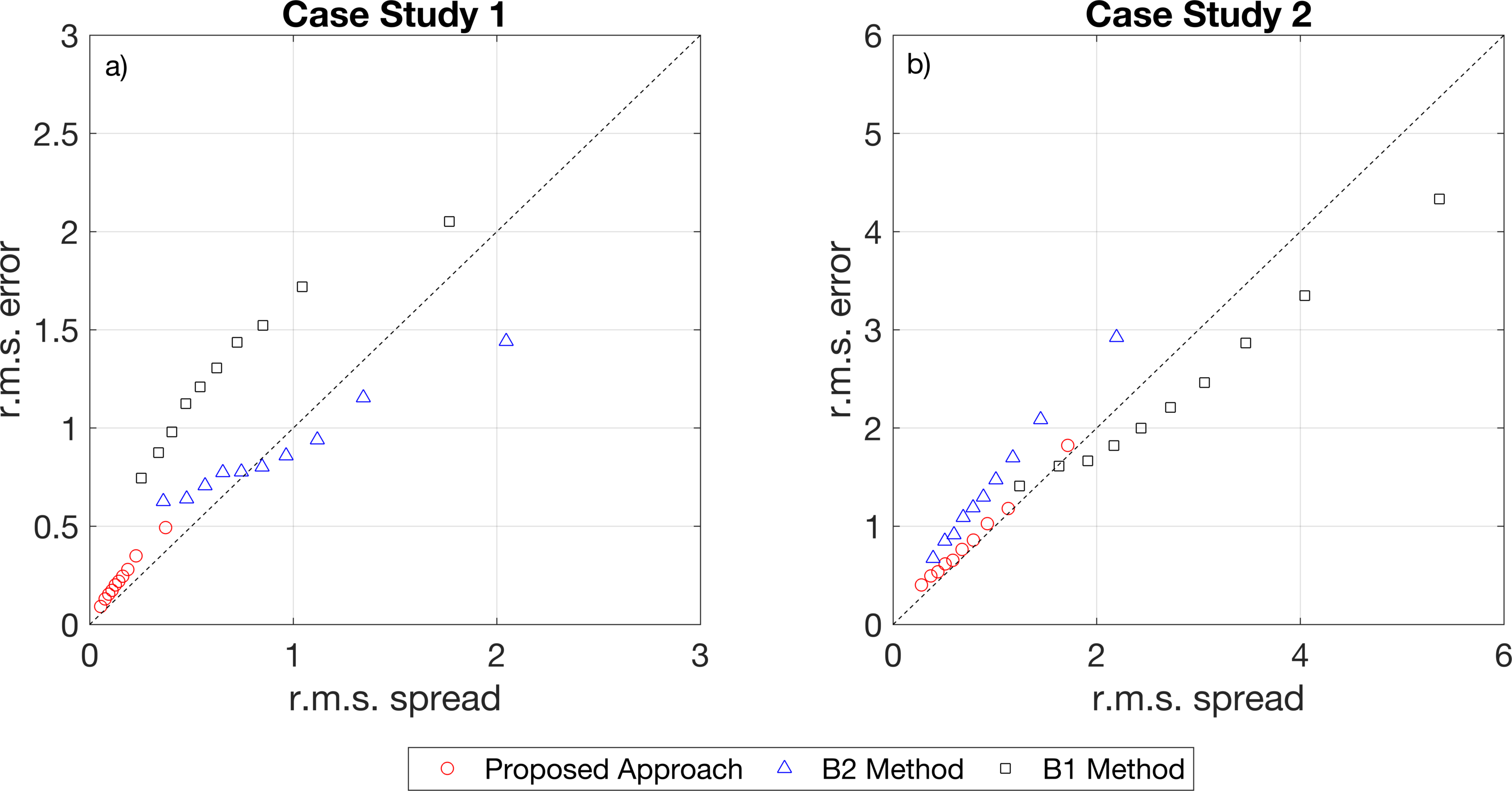}  
	\caption{Forecast r.m.s error vs r.m.s spread for both case studies as a measure of statistical consistency.  Results are provided for a forecast lead time of 0.6 MTU ($\approx$ 3 days). } 
	\label{fig:rmsespread}
\end{figure}

Overall, the proposed method was found to outperform the benchmark methods in all forecast metrics considered across a range of lead times, in both case studies.  This is demonstrated in Figure \ref{fig:rawscores} which shows the space and time averaged forecast score against lead time.  Forecasts from the proposed approach have better reliability, resolution and accuracy scores than the benchmarks, and are significantly more skilful at longer lead times (e.g. 0.6 MTU, or approximately 3 days).  The observed improvements are robust to different dynamical regimes, as indicated in Figure \ref{fig:skillscores} which shows the mean and standard deviation of skill scores computed over the 30 independent simulations.  Relative improvements are greatest when comparing to Method B1, where the proposed approach offers a 70\% improvement on average based on the RMSE and CRPS, although a sizeable improvement of 30\% is still apparent when comparing to Method B2 in Case Study 2 (see Figure \ref{fig:skillscores}).  
Forecast ensembles from the proposed approach also have better consistency properties, as shown by the RMS Error vs RMS Spread diagnostic plots (Figure \ref{fig:rmsespread}) where the points lie closer to the diagonal in the proposed approach.\\   

\section{Conclusions}
\label{S:5}
Characterizing model error is critical to ensure ensemble Data Assimilation methods produce high quality forecasts and analyses.  Accounting for model errors due to unresolved scales is particularly of interest in weather and climate modelling.  Numerous stochastic parameterization methods have been proposed for this purpose, although such methods generally rely on data or knowledge of the sub-grid scale processes and/or require observations of all resolved variables.  We develop a method that is suited to the more realistic condition where the resolved variables are only partially observed and knowledge of the sub-grid processes is unavailable. It allows for the estimation of complex error structures which depend on known covariates (e.g. state); requires no assumptions or specification or a parametric error distribution (e.g. Gaussian errors); considers the full range of statistical moments (not just bias and covariance); and avoids the need for numerical tuning typical of inflation and localization methods.  

The efficacy of the method is demonstrated through numerical experiments on the multi-scale Lorenz 96 model.  Comparisons are made to two existing methods that use data assimilation to estimate model errors offline, as these are amenable to the partially observed setting: 1) where the errors are assumed to be Gaussian with mean and covariance estimated from a sample of analysis increments; and 2) where model errors are estimated using long window weak constraint 4D-Var. The proposed approach is shown to recover model errors more precisely than the benchmark methods, thereby making it a more effective parameterization of the sub-grid processes.  It is also particularly useful for cases with highly non-Gaussian errors, as considered in this study.  Assimilation experiments with the ETKF show that the proposed approach leads to improved forecasts in terms of accuracy, reliability, resolution and consistency.  The conditional sum of squares minimization procedure in the proposed method also allows complex error structures to be estimated more precisely than with the least squares type 4D-Var approach.  The advantages of accounting for complex state dependent error relationships are also clearly demonstrated by the considerably poorer performance of the constant mean and covariance Gaussian error method. \\   

The proposed method is suited to multi-scale systems where a locality and 
homogeneity assumption can be made, i.e. where errors are influenced by neighbouring states instead of the full state vector and the error statistics are the same at each location in space or in parts of the state space with similar dynamics.  These assumptions help regularize the ill-posed problem of estimating model errors from partial observations.  Future work will investigate systems where such assumptions are inapplicable, although it is expected that other simplifying assumptions would be needed.  Finally, the method was applied to a case with negligible observation error, with some preliminary work including more prominent observation error.  Subsequent work will consider the more complex case of estimating model errors from noisy observations.

\section*{Acknowledgements}
 The research of S.Pathiraja has been partially funded by Deutsche Forschungsgemeinschaft (DFG)
- SFB1294/1 - 318763901 and by the UNSW Faculty of Engineering Postdoctoral Writing Fellowship.  PJvL acknowledges support from the EU-funded ERC grant CUNDA under number 694509.  We gratefully acknowledge Professor G. Gottwald and Professor S. Reich for insightful discussions during the development of this work.






\bibliographystyle{agsm}
\bibliography{modelerrorrefs.bib}







\end{document}